\documentclass[fleqn,usenatbib]{mnras}
\usepackage{newtxtext,newtxmath, bm}
\usepackage[T1]{fontenc}
\DeclareRobustCommand{\VAN}[3]{#2}
\let\VANthebibliography\thebibliography
\def\thebibliography{\DeclareRobustCommand{\VAN}[3]{##3}\VANthebibliography}
\usepackage{graphicx}	% Including figure files
\usepackage{amsmath}	% Advanced maths commands
\usepackage[dvipsnames]{xcolor}

\newcommand{\msun}{M_\odot}
\newcommand{\ta}{\tilde{a}}

\newcommand{\sr}{{superradiance }}
\newcommand{\srI}{{superradiant instability }}

\newcommand{\bh}{black hole }
\newcommand{\bhs}{black holes }

\newcommand{\ita}{\textit}
\newcommand{\mrm}{\mathrm}
\newcommand{\sra}{SR-active }

\title[Effects of Superradiance in AGNs]{Effects of Superradiance in  Active Galactic Nuclei}

\author[P. Sarmah et al.]{
Priyanka Sarmah$^{1,2}$\thanks{E-mail: sarmahpriyanka07@gmail.com}, Himanshu Verma$^{3,4}$\thanks{E-mail: verma.himanshu002@gmail.com}, Kingman Cheung$^{1,2,5}$\thanks{E-mail: cheung@phys.nthu.edu.tw},
and Joseph Silk$^{6,7,8}$\thanks{E-mail: silk@iap.fr}
\\
$^{1}$Department of Physics, National Tsing Hua University, Hsinchu 30013, Taiwan\\
$^{2}$Center for Theory and Computation, National Tsing Hua University, Hsinchu 30013, Taiwan\\
$^{3}$Department of Physics, Indian Institute of Technology Bombay, Powai, Mumbai, Maharashtra, 400076, India\\
$^{4}$Department of Physics and Astronomy, Louisiana State University, Baton Rouge, LA 70803, USA\\
$^{5}$Division of Quantum Phases and Devices, School of Physics, Konkuk University, Seoul 143-701, Republic of Korea\\
$^{6}$Institut d’Astrophysique de Paris (UMR7095: CNRS \& UPMC- Sorbonne Universities), F-75014, Paris, France\\
$^{7}$William H. Miller III Department of Physics and Astronomy, The Johns Hopkins University, Baltimore, MD 21218, USA\\
$^{8}$BIPAC, Department of Physics, University of Oxford, Keble Road, Oxford OX1 3RH, UK
}

\date{Accepted XXX. Received YYY; in original form ZZZ}

\pubyear{2025}
\begin{document}
\label{firstpage}
\pagerange{\pageref{firstpage}--\pageref{lastpage}}
\maketitle

\begin{abstract}
A supermassive black hole (SMBH) at the core of an active galactic nucleus (AGN) provides room for the elusive ultra-light scalar particles (ULSP) to be produced through a phenomenon called \textit{superradiance}. This phenomenon produces a cloud of scalar particles around the black hole by draining its spin angular momentum. In this work, we present a study of the superradiant instability due to a scalar field in the vicinity of the central SMBH in an AGN. We begin by showing that the time-evolution of the gravitational coupling $\alpha$ in a realistic ambiance created by the accretion disk around the SMBH in AGN leads to interesting consequences such as the amplified growth of the scalar cloud, enhancement of the gravitational wave emission rate, and appearance of higher modes of superradiance within the age of the Universe. We then explore the consequence of superradiance on the characteristics of the AGN. Using the Novikov-Thorne model for an accretion disk, we divide the full spectrum into three wavelength bands- X-ray ($10^{-4}-10^{-2}~\mu$m), UV (0.010-0.4~$\mu$m), and Vis-IR (0.4-100~$\mu$m) and observe sudden drops in the time-variations of the luminosities across these bands and Eddington ratio ($f_{\textrm{Edd}}$) with a characteristic timescale of superradiance. Using a uniform distribution of spin and mass of the SMBHs in AGNs, we demonstrate the appearance of depleted regions and accumulations along the boundaries of these regions in the planes of different band-luminosities and $f_{\textrm{Edd}}$. Finally, we discuss some possible signatures of superradiance that can be drawn from the observed time-variation of the AGN luminosities.
\end{abstract}

\begin{keywords}
accretion, accretion discs -- astroparticle physics -- instabilities -- quasars: supermassive black holes
\end{keywords}

%%%%%%%%%%%%%%%%%%%%%%%%%%%%%%%%%%%%%%%%%%%%%%%%%%

%%%%%%%%%%%%%%%%% BODY OF PAPER %%%%%%%%%%%%%%%%%%

\section{Introduction}
An active galactic nucleus (AGN) serves as a pivotal laboratory for fundamental particle physics due to the emission of high-energy particles and radiation originating from the proximity of a supermassive black hole (SMBH) at its core. Interestingly, the proximity of an SMBH is also a potential place to look for ultra-light scalar particles (ULSPs). The existence of such particles with mass possibly in the range $10^{-33}$-$10^{-10}$~eV is predicted in various theoretical~\citep{ulb1PhysRevD.16.1791, ulb2PhysRevLett.40.223, ulb3PRESKILL1983127, ulb4PhysRevD.91.015015, ulb5Chikashige:1980ui} and observational~\citep{ulb6Bar:2018acw, ulb7Hlozek:2014lca, ulb8Khmelnitsky:2013lxt, ulb9Irsic:2017yje} contexts. The ULSPs can be produced via the instability of the corresponding scalar field perturbation around a spinning black hole through a process called \textit{superradiance}~\citep{sr00Brito:2015oca}. As a result of this intriguing process, a cloud of scalars grows at the expense of the mass and spin angular momentum of the black hole (BH). It can only occur when the BH is spinning faster than a critical spin.

The rate of the superradiant growth depends on the spin of the BH and the gravitational coupling constant $\alpha = M \mu$, where $\alpha$ is a measure of relative size of the BH horizon with respect to the Compton wavelength of the scalar, $M$ is the mass of the BH, and $\mu$ is the mass of the scalar~\citep{sr00Brito:2015oca}. Particularly, for the cloud to grow within the current age of the universe, the gravitational coupling $\alpha$ should approximately lie within 0.01 to 1 for an extremally spinning SMBH.  Hence, one can get the maximum possible mass range of an SMBH to be  $M\in [10^{7} \msun, 10^{9} \msun] 10^{-19} \mrm{eV}/\mu$, for which the cloud growth could be fast enough to grow within the age of the Universe. This further implies that the scalars lying in the range of mass $10^{-20}$-$10^{-16}$~eV can be produced in the vicinity of SMBHs ($10^6-10^{10}~M_\odot$) within observable time-scales. The presence of the scalar cloud around an SMBH can be tested in various observational scenarios. These include the studies of scalar clouds affecting the black hole images~\citep{Davoudiasl:2019nlo, Roy:2019esk, 2022arXiv220803530S, Chen:2021lvo, Chen:2022oad, Wang:2023eip, sr19Roy:2021uye, Chen:2022kzv, Chen:2022nbb}, gravitational wave (GW) emission from the scalar cloud~\citep{sr24Brito:2017wnc, Hannuksela:2018izj, sr5Yoshino:2013ofa, sr11Arvanitaki:2014wva, sr4Baryakhtar:2020gao, sr10Guo:2022mpr}, scalar clouds perturbing the orbit of the S2 star~\citep{Cardoso:2011xi, Ferreira:2017pth, GRAVITY:2019tuf, GRAVITY:2023cjt}, and scalar clouds decaying to photons due to scalar-photon coupling~\citep{Brito:2014nja, Yoo:2021kyv}. 

Another profound prediction of the superradiant instability is the existence of a \textit{depletion region} in the spin versus mass plane (we will refer to this as the Regge plane) of BHs~\citep{sr20Arvanitaki:2010sy}. A superradiance time exists until when a BH can stay in the depletion region after which it loses its spin and settles down at the critical spin which depends on the BH mass. This forms a \textit{critical spin curve} on the Regge plane. The SMBHs, therefore, with ages larger than the \sr time-scale should never stay within the depletion region but rather should cluster along the critical spin curve. Hence, the observation of the depleted region and the clustering along the critical spin curve of SMBHs can provide another smoking gun signature of ULSPs in the Universe~\citep{sr1Brito:2014wla, sr21Stott:2018opm, sr12Cardoso:2018tly, sr22Unal:2020jiy}. Such observations however require a precise measurement of the spin of the SMBHs which in the most realistic scenario possible is when the SMBH is surrounded by a luminous accretion disk. In that case,  methods like the use of reflection spectrum and the continuum fit~\citep{method1Reynolds:2019uxi} can provide precise measurements of the spin but they require very high signal-to-noise (SNR) ratio data. On the other hand, there are black hole imaging techniques (such as Event Horizon Telescope) to measure the spin of the SMBHs~\citep{method2EventHorizonTelescope:2019pgp} but applicable to the most nearby AGNs only. 

In all the practical cases when the spin and mass measurements are possible, the SMBHs are not isolated but rather they go through various accretion and merger phases at the galactic center~\citep{2012MNRAS.419.2797F, 2014MNRAS.442.2304H}. These phases, particularly the accretion process, can alter the dynamics of the SMBHs on the Regge plane quite significantly by dragging them into the depletion region~\citep{sr9Hui:2022sri}. Reference~\citep{sr1Brito:2014wla} found that the depletion region can shrink in the presence of accretion. Hence, in reality, the depletion region can occasionally be populated even in the presence of ULSPs in the universe. To enhance the likelihood of finding ULSP signatures on the Regge plane, it would therefore require large statistics of spin and mass measurements of SMBHs. On the other hand, a relatively larger population of AGNs ($\sim 500,000$) have so far been detected by the Sloan Digital Sky Survey (SDSS) whose luminosities in various bands and spectra have been measured~\citep{eBOSS:2017qtj, 2020ApJS..249...17R}. The statistics and the precision of the measured characteristics of AGNs are expected to improve further with upcoming data releases from telescopes like The Dark Energy Spectroscopic Instrument (DESI)~\citep{DESI} and eventually from the Large Synoptic Survey Telescope (LSST)~\citep{LSST:2008ijt}. This very fact motivates us to consider the distribution of AGN characteristics such as luminosities as a potential probe for the ULSPs and its possible complementarity with the Regge plane in the absence of precise spin measurements of SMBHs.

The luminosity of AGN has a well-known maximum limit called the Eddington luminosity that depends on the mass of the SMBH as  $L_\textrm{Edd}\approx 1.26\times10^{38}~\textrm{erg/s}~M/M_\odot$. Now in the presence of superradiant activity, the mass and spin of the SMBH will vary in a certain way leading to a time-variation in the flux emitted from the accretion disk and hence in the actual bolometric luminosity which are functions of the BH spin and mass. Therefore, for AGNs that can potentially have the signature of \sr due to a scalar of mass $\mu$, their corresponding Eddington luminosity will approximately lie within
\begin{equation}
\label{eq:LEddrange}
    L_\textrm{Edd}\in [10^{45} , 10^{47}]~\textrm{erg/s} \frac{10^{-19} \mrm{eV}}{\mu}.
\end{equation}

This fact constitutes an important part of this work where we attempt to calculate the superradiance-induced time variation and the distribution of the AGN luminosities in various bands and the Eddington ratio. This work in general presents a detailed study of the superradiant activity in the SMBH at the core of AGN. We divide the study into three parts-
\begin{itemize}
\item We start by discussing the role of accretion on the dynamics of the BH-cloud system that passes through multiple modes of superradiance. Through simulations, we highlight the effect of accretion on the gravitational coupling evolution which controls the cloud growth and power of gravitational wave emission from the cloud. 
\item We then aim to calculate the time variation of the Eddington ratio and luminosities of AGN in various wavelength bands using the Navikov-Thorne (NT) model of accretion disks while the accreting SMBH at its core goes through superradiant instability.
\item We then evolve a uniformly distributed accreting SMBHs on the Regge plane in the presence of a scalar field and derive the superradiance-modified distribution of corresponding AGNs in various planes of X-ray, UV, and Vis-IR wavelength bands and Eddington ratio.
\end{itemize}

We achieve the above by first following the simple approach of numerically solving the differential equations that govern the dynamics of the BH-cloud system in the presence of accretion. Solving these equations under the quasi-adiabatic approximation~\citep{sr1Brito:2014wla}, we observe an enhanced growth of the scalar cloud mass due to accretion (up to 25\% of the black hole mass) compared to the usual expectation of  10\%~\citep{Herdeiro:2021znw}, also known as over-threshold-superradiance (or over-superradiance)~\citep{sr9Hui:2022sri}. We demonstrate about eight orders of enhancement in the GW emission rate in the case of SMBH accreting at the rate of half of the Eddington luminosity. This enhanced GW emission leads to the cloud's fate being primarily depleted by the GW emission from the cloud. Also, it is interesting to see the appearance of the higher order modes of \sr which is otherwise not feasible without accretion.
  
We then calculate the effect of the superradiant spin-down on the accretion disk characteristics around the black hole in an AGN.  Given the time evolution of the spin and mass of the BH obtained from the superradiant growth, we calculate the spin-dependent flux coming from the accretion disk using the NT model. The color-corrected continuum spectrum of the AGN is then calculated using the flux profile obtained from the NT model following the ref.~\citep{2011MNRAS.414.1183K}. We divide the full spectrum into three distinct wavelength bands- X-ray ($10^{-4}-10^{-2}~\mu$m), UV (0.010-0.4 $\mu$m), and Vis-IR (0.4-100$~\mu$m) and observe that the spin-down effect alters the X-ray and UV luminosity most remarkably. We further obtain the spin-down effect on the Eddington ratio. For a given accretion rate, the Eddington ratio is controlled by the radiation efficiency which is a function of the BH spin. Hence because of superradiant spin-down, the Eddington ratio would be suppressed for a long time before reaching a maximal value. It is interesting to observe this feature which is in contrast with the case of no \srI where the Eddington ratio is expected to grow monotonically.

%%%%%% 6- talk about distribution %%%%%%%%%%%%%%%
We finally derive the impact of the characteristics seen in the Regge plane on the distribution of the AGN band-luminosities. We start with the time evolution of $10^4$ black holes uniformly distributed on the Regge plane and notice the appearance of the depletion region and accumulation along specific boundaries in the plane depending upon the scalar mass. We then calculate the  AGN characteristics of these $10^4$ SMBHs using the above description. We observe similar depletion regions on the planes of various band-luminosities and  Eddington ratios. The salient feature of these distributions is that the AGN luminosities tend to align along the various tracks on these planes where these tracks correspond to the critical-spin curve on the Regge plane for various modes of superradiance that are taken into account.
%%%%%%%% signatures and future work%%%%%

We conclude this work by pointing out a few possible implications of our proposed superradiance-induced time-varying luminosity in the form of galactic outflows around AGN  and the Lyman-$\alpha$ forest from bright quasars. The layout of the paper is as follows- we start with the main aspects of the \sr relevant for our work in subsection \ref{sec:sr_basics}, followed by which we introduce and elaborate the case of \ita{superradiant-active or SR-active} \bh in subsection \ref{subsec:SRactive}. Then in the following section \ref{sec:srwithacc}, we discuss important findings of the effects of accretion on \sr evolution. Starting with the description of the accretion model in \ref{subsec:accretionmodel}, we highlight the important features of the evolution of the scalar cloud and spin of the accreting \sra \bhs in \ref{subsec:evolutioneqns}. In \ref{subsec:Lumandfedd}, we discuss the effects of \srI on the Luminosity in different wavelength bands and Eddington ratios for a specific case of the scalar and black hole mass. In section \ref{sec:observables}, we show the possible observable effects on the distribution of the luminosity and Eddington ratio for \sra BHs. Finally, we discuss and summarize in section \ref{sec:conclusion} with a mention of possible signatures of the spin-down effects in AGN.

\section{Superradiance of Supermassive Black Holes}
In this section, we will lay out the basic framework of \sr required to study the spin-down of an SMBH in isolation i.e. without any accretion. We will then discuss the time-scale associated with the phenomenon and the relevant parameter space of SMBHs for a given scalar mass. 

\subsection{Basics of superradiance}\label{sec:sr_basics}
Superradiance, in general, is a process where waves of massive bosonic particles scattered off a rotating black hole extract its energy and angular momentum under favorable conditions. The space-time nature around a black hole of mass $M$ and spin $a$ can be explained by the \textit{Kerr metric} given in Boyer-Lindquist coordinates, $(t, r, \theta, \phi)$, as  
\begin{align*}
d s^2 &=-\left(1-\frac{2 M r}{\rho^2}\right) d t^2-\frac{4 M r a}{\rho^2} \sin ^2 \theta d t d \phi +\frac{\rho^2}{\Delta} d r^2 +\rho^2 d \theta^2+\\
&\left(r^2+a^2+\frac{2 M r a^2 \sin ^2 \theta}{\rho^2}\right) \sin ^2 \theta d \phi^2,
\end{align*}
where
$\rho^2=r^2+a^2 \cos ^2 \theta \text { and } \Delta=r^2+a^2-2 M r $.
Here spin $a $ of the black hole is given in the units of $M$, i.e. $a \equiv J / M$ where $J$ is the spin angular momentum of the black hole. We will use the notation of the dimensionless \ita{spin parameter} defined as $\ta \equiv a/M$. The spin parameter $\tilde{a}$ can take any value between 0 to 1, where $\tilde{a}=0$ corresponds to the Schwarzschild black hole. 

The superradiance process around a Kerr BH requires an initial seed scalar field (or any bosonic field, in general) to begin the amplification. This seed perturbation can arise from quantum fluctuations in the scalar field, which would exist near the BH if such a field is present in our universe. The superradiant instability of these perturbations around the Kerr black hole can then cause the field to grow, eventually forming a cloud of scalar particles around the black hole. The condition for which the instability occurs is~\citep{sr00Brito:2015oca}
\begin{eqnarray}
    \label{eqn_srcondition}
    \omega_R < m\Omega,
\end{eqnarray}
where $\omega_R$ is the real part of the angular frequency $\bm{\omega}$ of the scalar field. The scalar field, $\Phi$ being the solution of the  Klein-Gordon equation ($\Box\Phi+ \mu^2\Phi=0$), can be decomposed into  $\Phi=e^{-i \bm{\omega} t}R_{nlm}(r)S_{lm}(\theta, \phi)$ around the black hole, where $S_{lm}(\theta, \phi)$ are the spheroidal harmonics and the radial part $R_{nlm}(r)$ can be obtained by solving an eigenvalue problem obtained by the Klein-Gordon equation. Solving the eigenvalue problem under the appropriate boundary conditions leads to quasi-bound states with quantum numbers $n,l,m $ and energy eigenvalues $\bm{\omega}$, very similar to the Hydrogen atom solutions.  The scalar cloud grows in each mode $nlm$ draining the black hole's angular momentum as soon as the condition in  eq.~\ref{eqn_srcondition} is satisfied \ita{i.e.} when BH's angular velocity $\Omega=\frac{a}{2M r_{+}}$  at the event horizon ($r_{ \pm}=M \pm \sqrt{M^2-a^2}$) exceeds the particle's angular velocity $\omega_R$.

Interesting physics of \srI is derived when its rate is the largest, and it happens for the case when the Compton wavelength ($\lambda=1/\mu$) of the scalar particle of mass $\mu$ is comparable to the horizon size of the BH($r_g\equiv M $). This condition can be re-expressed further in terms of the parameter $\alpha$, known as the \ita{gravitational fine structure constant} and is given in terms of Planck unit $G=c=\hbar=1$ as~\citep{Ferraz:2020zgi}
 \begin{equation}\label{srcondition}
       \alpha \sim \left(\frac{\mu}{10^{-10} \mathrm{eV}}\right)\left(\frac{M}{M_{\odot}}\right)\lesssim 1 .
\end{equation}
Now,  in the small  $\alpha$ limit, the energy eigenvalue solutions $\bm{\omega}$($\equiv \omega_R+i \omega_I$) exhibit a small positive imaginary part resulting in the superradiant growth in a particular mode $nlm$. The rate $\omega_I^{nlm}$,  at which the growth occurs is given by  the imaginary part of the solution~\citep{sr27Baumann:2018vus}
\begin{equation}\label{eqn_srrate}
    \omega_I^{nlm}\equiv \alpha^{4 l+5} \frac{r_{+}}{M}(m \Omega-\omega_R) C_{n l m},
\end{equation}
where 
\begin{align*}    
    C_{n l m}=\frac{2^{4 l+1}(2 l+n+1) !}{(l+n+1)^{2 l+4}n!}\left(\frac{l !}{(2 l) !(2 l+1) !}\right)^2 
\times \prod_{j=1}^l[j^2(1-a^2/r_g^2)&\\+4 r_{+}^2(m \Omega-\omega_R)^2].
\end{align*}
This analytical expression for the rate deviates from the numerical rate given in \citep{Ferraz:2020zgi} for a higher spin and $\alpha \geq 0.1$. Since the analytical rate is lower than the numerical rate for higher spin and higher alpha, our prediction for the signal is therefore underestimated in that limit.

The real part of solution  $\bm{\omega}$,  denoted  as $\omega_{R}$, is given, to second order in $\alpha$, by
\begin{equation}\label{eqn_wr}
    \omega_{R}^{nlm} \approx \mu\left(1-\frac{\alpha^2}{2(n+l+1)^2}\right).
\end{equation}
Under the \sr condition in eq.~\ref{eqn_srcondition}, the growth rate $\omega_I^{nlm}$ for a particular mode $nlm$ becomes positive indicating an exponentially large population ($N_{\mrm{max}}$) of the scalars around the BH. During this growth, the black hole loses its spin angular momentum and continues to lose  until its spin $\tilde{a}$ reaches a critical value $ \tilde{a}_{\mrm{crit}}$
\begin{align}\label{eq:acrit}
    \tilde{a}_{\mrm{crit}}(M,\mu,m)&\approx \frac{4 m M \omega_R}{m^2+4\omega_R^2M^2},
\end{align}
below which the condition of \sr will no longer be satisfied. This conveys the fact that in the presence of a given scalar, it will be most likely to find a BH of mass $M$  at the critical spin value given by $\tilde{a}_{\mrm{crit}}$.

From the \sr rate $\omega_I^{nlm}$,  one can now estimate the instability time scale $\tau_{nlm}$ for a given scalar mass $\mu$ as
\begin{eqnarray}
    \tau_{nlm} \equiv \frac{1}{2\omega_I^{nlm}}.
\end{eqnarray}
Furthermore, as the \srI develops, two important back-reactions (non-linear effects) of the scalars come into play -- the GW emission and non-linear self-interaction of scalars in the cloud. The effect of the latter that arises due to the higher order terms in the axion sine-Gordon potential, is extensively studied in Ref.~\citep{sr4Baryakhtar:2020gao}.  In our work, we assume no self-interaction of the scalar field and simply focus on the dynamics of the cloud under GW emission. The scalars in the cloud can annihilate to produce gravitons and emit gravitational waves of frequency $f \sim 2\mu \approx 5\times10^{-5}~\mrm{Hz} (\frac{\mu}{10^{-19}\mrm{eV}})$. Due to this, the cloud loses its energy as well as angular momentum in the form of GWs. The rate of energy emitted in terms of GWs can be obtained from the scalar stress-energy tensor and is given in the flat space-time approximation by~\citep{sr5Yoshino:2013ofa}
\begin{align}\label{eqn_Egwrate}
\dot{E}_\mrm{GW}^{nlm} \equiv \frac{d E_{\mrm{GW}}^{nlm}}{d t}=C_{n \ell}\left(\frac{M_s}{M}\right)^2(\mu M)^{4 \ell+10},
\end{align}
where $C_{n \ell}=\frac{16^{\ell+1} \ell(2 \ell-1) \Gamma(2 \ell-1)^2 \Gamma(\ell+n+1)^2}{n^{4 \ell+8}(\ell+1) \Gamma(\ell+1)^4 \Gamma(4 \ell+3) \Gamma(n-\ell)^2}$.
The Schwarzschild metric background gives a different prefactor $C_{n \ell}$\citep{sr00Brito:2015oca}. The percentage change in  $\dot{E}_{GW}$ could be as large as 100\% compared to the flat metric assumption for the particular benchmark point of BH and scalar mass discussed here. Such enhancement cannot alter the superradiant growth rate, and hence, there will be no significant effect on the spin evolution as far as high accreting systems like AGNs are concerned. Thus, the prediction of the depletion region and its evolution in the spin vs. mass (shown in sec.~\ref{subsec:reggeplane}), as well as in different planes of band luminosities of the AGNs (shown in sec.~\ref{subsec:luminvariousbands}), will remain intact. However, the change in GW emission rates due to a realistic metric could have a more substantial impact on spin evolution if the accretion rate is lower than the accretion of the superradiant cloud during the collapse phase driven by GW emission.
Compared to the \srI rate $\omega_I^{nlm}$, the GW emission rate is suppressed in the limit $\alpha<1$ for an isolated black hole. A detailed discussion of the time-scale involved in the \srI is presented in the following section. 

\subsection{Superradiant active SMBHs}
\label{subsec:SRactive}
\begin{table}
\centering
\begin{tabular}{|p{0.7cm}|p{6.8cm}|}
 \hline
 modes & SR-active black hole mass range\\
 \hline
 011 & $M \in \left[4.2\times10^7~M_\odot \left(\frac{10^{-19}~\textrm{eV}}{\mu}\right)^{1/9},~6.7\times10^8~M_\odot\frac{10^{-19}~\textrm{eV}}{\mu} \right]$\\
 & \\
 022 & $M \in \left[2.0\times10^8~M_\odot \left(\frac{10^{-19}~\textrm{eV}}{\mu}\right)^{1/13},~1.3\times10^9~M_\odot\frac{10^{-19}~\textrm{eV}}{\mu} \right]$\\
 & \\
 033 & $M \in \left[4.8\times10^8~M_\odot \left(\frac{10^{-19}~\textrm{eV}}{\mu}\right)^{1/17},~2.0\times10^9~M_\odot\frac{10^{-19}~\textrm{eV}}{\mu} \right]$\\
 \hline
 \end{tabular}
\caption{SR active mass ranges of extremally spinning SMBHs in the dominant modes $nlm=011,022,033$ corresponding to different scalar mass $\mu$.} 
 \label{table_1}
 \end{table}

We confine our study to the parameter space of SMBHs which can spin down due to \sr within the age of the universe ($T_\textrm{univ}\sim 10^{10}$ years) and we term them as \textit{superradiant active} (SR-active) SMBHs. We also restrict our study to the three most dominant modes $nlm=011$, $022$, $033$ of the scalar field perturbation around the SR-active SMBHs. For \sra SMBHs, there exists a range of mass of the \bhs that depends on the scalar mass and a minimum spin  $\tilde{a}_{\mrm{crit}}$ above which these SMBHs will be able to spin down within the observable time of the universe. The lower mass limit can be estimated using $\tau_{nlm}=T_\textrm{univ}$ and by approximating $\omega^{nlm}_I\approx C_{nlm} \alpha^{4l+5} \frac{m}{2M}$ for extremal spin. The upper mass limit can be estimated when the growth rate for the extremal spin vanishes i.e. $m\Omega \approx \mu$. Thus, the parameter region of SR-active black holes is given by
\begin{equation}
    M \in \frac{1}{\mu} \left[\left(\frac{1}{m C_{nlm}T_\textrm{univ} \mu}\right)^{\frac{1}{4l+4}},~\frac{m}{2}\right]~\cup \tilde{a}>\frac{4 m \alpha}{m^2+4\alpha^2},
\end{equation}
where $C_{nlm}$ approximates to 1/48, 1/13882, and 1/12817535 corresponding to 011, 022, and 033 modes respectively and $\omega_R\approx \mu$. In Table~\ref{table_1}, we explicitly give the mass range of the SR-active black hole for each of the three modes. For a scalar mass $\mu$, black holes lying outside of the SR-active mass range of the given mode will not be able to grow the superradiant scalar cloud within an observable time and hence will be SR inactive for the mode. For BH with spin less than $\tilde{a}<0.998$, the relevant mass range will be a subset of the mentioned ranges.

\begin{figure*}
\centering  
\includegraphics[width=0.49\textwidth]{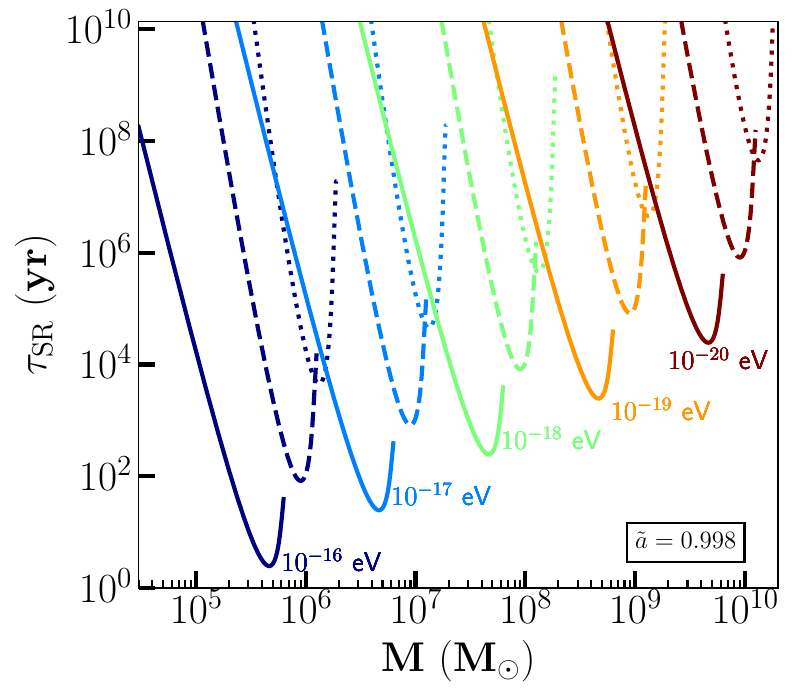}
\includegraphics[width=0.49\textwidth]{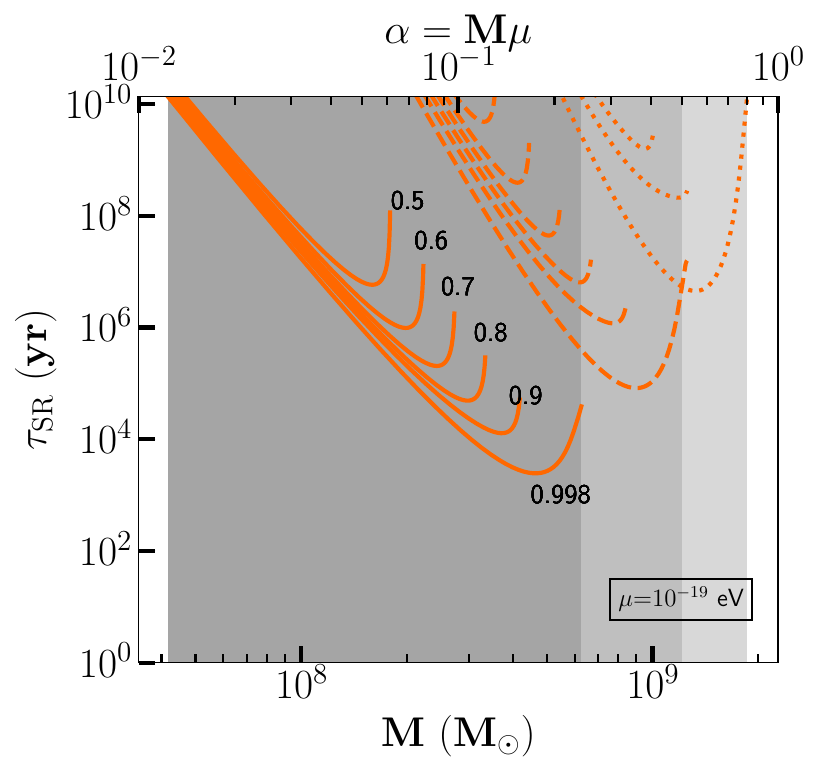}
\caption{This figure shows the dependence of the \sr timescale $\tau_\textrm{SR}\equiv \tau_{nlm}$ on the black hole parameters and the scalar mass $\mu$ for three most dominant modes $nlm=011,~022$, and $033$. In the \textit{left plot}, we show the variation of $\tau_\textrm{SR}$ with the black hole mass keeping the spin parameter at the extremal value $\tilde{a}=0.998$ and different values of scalar mass $\mu$~(in eV) are labeled. The modes $nlm=011$ (solid), $nlm=022$ (dashed), and $nlm=033$ (dotted) are shown with different lifestyles. Time for \srI is the lowest for the lowest angular momentum state ($l=m=1$) which stands out as the most dominant out of the three. The \textit{right plot} shows the \sr time-scale variation with black hole mass for $\mu=10^{-19}$~eV and different curves are drawn for various spin values as annotated in the figure. The whole SR-active mass range is split into three parts dominated by 011, 022, and 033 modes as shown by different gray-shaded regions.}
  \label{fig:Tau_sr}
\end{figure*}

In fig.~\ref{fig:Tau_sr}, we discuss the SR time-scale ($\tau_{nlm}$) in which there will be an exponential growth of the cloud around an extremally spinning SMBH with mass falling in the range 10$^4$-10$^{10}$ $\msun$. The main purpose of the figure is to give an idea about the timescale of \sr and the relevant mass range of \sra black holes corresponding to a certain scalar mass and spin. Different line styles represent the three dominant ($nlm=011, 022, 033$) modes and various colors show the scalar masses, $\mu$.  For example, the presence of a scalar mass $\mu=10^{-19}$ eV (shown in yellow) can induce the SMBHs roughly lying in the mass range of 7$\times10^7 - 2\times10^9$ $\msun$ to undergo \srI within the lifetime of the Universe. Initially, for a fixed scalar mass,  $\tau_{nlm}$ decreases with $M$ and touches the minimum, resulting in an exponential growth of scalar cloud around the BH  at the cost of BH spin angular momentum.  The cloud eventually ceases to grow as soon as the condition for \sr (\ref{eqn_srcondition}) is no longer satisfied. At this point, the spin of the black hole reduces to the critical value $\tilde{a}_{\mrm{crit}}$. This explanation for the time-scale holds for any mode $nlm$ of superradiance. However, as we can see, the time-scale is lowest for the fastest-growing angular momentum state with  $l=m=1$. This is evident from the rate eq.~\ref{eqn_srrate}, which decreases for higher angular momentum states ($l=2,3$) for $\alpha \leq 1$. The left plot of the figure thus provides an idea of the time-scale when the \bhs in a particular mass range will be superradiantly active because of the presence of a certain scalar particle. This time-scale increases as scalar mass decreases.

In the right panel of the figure, we show how the time-scale changes as we vary the spin parameter of the BH. We choose to display the case of scalar mass $\mu=10^{-19}$~eV. As it is apparent from the figure, the higher the spin, the lower will be the time-scale of the superradiance. Moreover, for each spin value, the full mass range of the SMBH  relevant for a given $\mu$, can be divided into three ranges where one of the three modes of \sr will dominate the most. As can be seen for the case $\mu=10^{-19}$~eV, for extremal spin SMBHs lying in the range of $4.2\times10^7 - 6.3\times10^8$~$\msun$ (dark grey-shaded region), the first mode of \sr ($nlm=011$) will be most dominant. For the range $6.3\times10^8 - 1.2\times10^9$~$\msun$ (lighter grey region), it is the second mode ($nlm=022$) that is predominant, whereas for the SMBH whose mass falls in $1.2\times10^9 - 1.9\times10^9$~$\msun$ (lightest grey-shaded region), the third mode ($nlm=033$) will prevail the most. It should be noted that the mass range in the shaded region in each of the modes represents the corresponding dominant mode, whereas the range given in table \ref{table_1} provides the maximum possible mass range of the BH that would show \sr effects within the observable time. 

Next, we will discuss the case of the accreting SMBH at the core of an AGN. The central SMBH in the AGN accretes matter from the accretion disk around it. The spiraling matter falling from the accretion disk increases the mass and also transfers the angular momentum to the BH. With the knowledge of the parametric dependence of the superradiant timescale from Fig.~\ref{fig:Tau_sr}, one can thus identify whether the central SMBH of an AGN is \sra or not and hence will be called such an AGN as SR-active too. Because of accretion that feeds mass to the BH, it may be possible for a BH whose initial mass is smaller than the minimum required mass of the  SR-active BH, to eventually fall in the SR-active mass range. A detailed discussion of the \sra  SMBH at the core of an AGN is what follows in our next section.

\section{Active Galactic Nucleus Evolving with scalar field}
\label{sec:srwithacc}
In this section, we study the time evolution of the accreting SMBH at the core of an AGN in the presence of an ultra-light scalar field in our universe. We will first discuss the basic accretion model and then derive the time-varying characteristics of the SR-active AGN.

\subsection{AGN and the accretion model of SMBH at the core}\label{subsec:accretionmodel}
An AGN consists of a disk of matter swirling around an SMBH at the core of a galaxy. The accretion disk dominantly feeds the SMBH as compared to the ambient gas at the galactic center. The viscosity and the velocity gradient across the gas layers in the disk increase towards the center of the BH, which leads to higher friction between the layers near the event horizon than the layers that are far away from the event horizon. Because of this increasing friction, there is a gradual loss of kinetic energy of the particles and hence the loss of angular momentum towards the BH. Eventually, with the losing angular momentum, the particles fall into the fate of the BH. 

As matter falls from the accretion disk, it increases the mass of the SMBH and increases (decreases) the spin angular momentum of the SMBH if the accretion disk is co-rotating (counter-rotating) with the SMBH. Material accelerated toward the SMBH from the accretion disk undergoes ionization, with their gravitational potential energy transforming into electromagnetic radiation, which fuels the AGN to become the most luminous object in the sky. The emitted radiation from an AGN can be so huge that its radiation pressure counteracts the influx of gases, establishing an upper limit to the AGN luminosity, known as the Eddington luminosity and given by~\citep{AGNatcosmicdawnGriffin}
\begin{equation}
\label{eq:LEdd}
    L_\textrm{Edd} = \frac{4\pi G M m_p c}{\sigma_T} \approx 1.26\times10^{38} \textrm{erg/s}\frac{M}{M_\odot},
\end{equation}
where $m_p$ represents the proton mass and $\sigma_T$ signifies the Thomson scattering cross-section of an electron with photon. In practice, the luminosity of an AGN can be characterized by a fraction of Eddington luminosity as $f_\textrm{Edd} \equiv L/L_\textrm{Edd}$, known as the Eddington ratio, where $L$ is the bolometric luminosity of an AGN.

\begin{figure}
 \center
\includegraphics[width=\columnwidth]{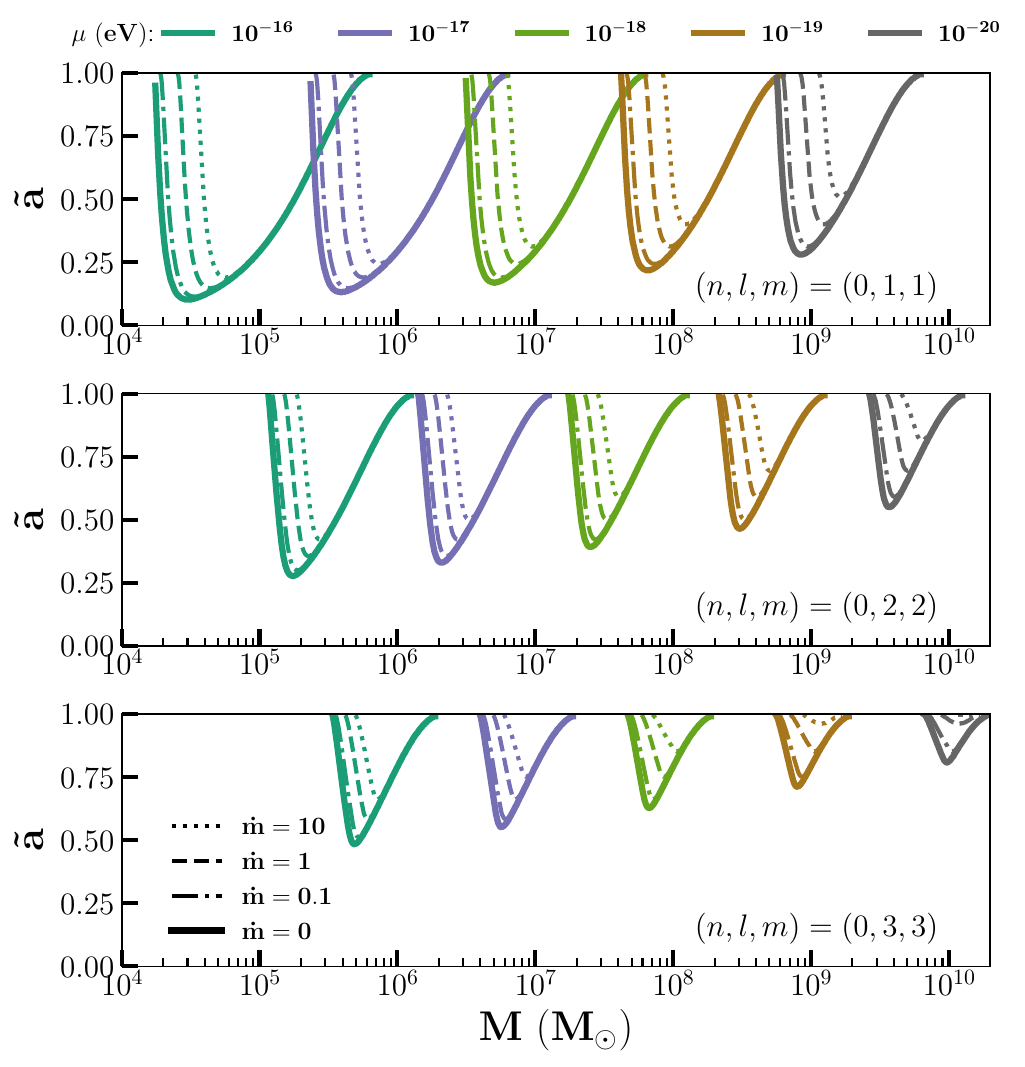}
  \caption{This figure depicts the relevant black hole parameter space ($\tilde{a}$, $M$) shown by the region above the curves, for which the growth of the scalar cloud can be significant and the region may get depleted. The regions are shown for the most dominant modes $nlm = 011, 022,$ and $033$ in the \textit{upper}, \textit{middle}, and \textit{lower} panels respectively. The regions above the solid-line curves are such that $\tau_{nlm}<T_\textrm{univ}$ where superradiance can occur within the age of the universe without accretion ($\dot{m}=0$). The region bounded by the broken-line curves are obtained using $\tau_{nlm}< \textrm{min}(\tau_\textrm{acc}, ~T_\textrm{univ})$ in which the superradiant spin-down rate will be larger than the spin-up rate, implying that the accreting black holes ($\dot{m}\neq0$) lying in this regions will undergo \sr first. This simply indicates that it is the smaller mass part of the depletion region that may get shrunk due to the presence of accretion, where the dimensionless accretion rate parameters $\dot{m}$ (accretion rate normalized to Eddington Luminosity) are taken to be $0.1,~1,$ and $10$.}
  \label{fig:Depletion}
\end{figure}

The amount of matter required to infall per unit time from the accretion disk to produce a given bolometric luminosity depends on the fraction of the rest mass energy of the accreting gas radiated away as electromagnetic radiation, also known as radiative efficiency ($\epsilon$). The swirling matter in the disk loses energy due to the viscosity till it reaches the inner edge of the accretion disk, which is typically taken to be the innermost stable circular orbit (ISCO), and then accreted by the black hole at the center. Therefore, the radiative efficiency can be estimated as the binding energy per unit mass of the matter at ISCO, estimated as ~\citep{2011MNRAS.410...53F},
\begin{equation}
    \epsilon(\ta) =1-e_\textrm{isco},
\end{equation}
where $e_\textrm{isco} =  \sqrt{1 - \frac{2}{3}\frac{M}{r_\textrm{isco}} } $, the radius of the ISCO is given by $r_\textrm{isco} = (G M/c^2)[3+Z_2- \ta_\textrm{sign}\sqrt{(3-Z_1)(3 + Z_1 + 2 Z_2)}]$, $Z_1 = 1+(1-\ta^2)^{1/3}[(1+\ta)^{1/3}+(1-\ta)^{1/3}]$, and $Z_2 = \sqrt{3 \ta^2 + Z_1^2}$. It should be noted that $\ta_\textrm{sign}$ is the sign of the spin, which is positive for co-rotating and negative for counter-rotating disk relative to the spin of the BH.  Assuming the rate of infalling matter from the disk, or simply the accretion rate is $\dot{M}_\textrm{disk}$, we can calculate the bolometric luminosity using the radiative efficiency as
\begin{equation}
    L = \epsilon(\ta) \dot{M}_\textrm{disk} c^2.
\end{equation}
It is convenient to parameterize the accretion rate relative to the $L_\textrm{Edd}$ as $\dot{m} \equiv \dot{M}_\textrm{disk}c^2/L_\textrm{Edd}$. We can therefore write the bolometric luminosity in terms of radiative efficiency, $\dot{m}$, and Eddington luminosity as  
\begin{align}
\label{eq:Lbol}
L=\epsilon(\ta) \dot{m} L_\textrm{Edd}.
\end{align}
With this, one can re-write the Eddington ratio $f_\textrm{Edd} \equiv L/L_\textrm{Edd}$ as
\begin{align}
\label{eq:fedd}
f_\textrm{Edd}=\epsilon(\ta) \dot{m}.
\end{align}

It should be noted that the accretion rate parameter $\dot{m}$ has to be large enough to form a disk which we consider to be above 0.01~\citep{Abramowicz:2011xu}. As the fraction of energy radiated away from the disk is $\epsilon(\ta) \dot{M}_\textrm{disk} c^2$, BH can accrete the remaining fraction $1-\epsilon$ of the infalling matter with the rate of 
\begin{align}\label{eqn_Macc}
\dot{M}_\textrm{acc} &= (1-\epsilon(\ta)) \dot{M}_\textrm{disk} = \frac{M}{\tau_\textrm{acc}},
\end{align}
where the accretion time scale is defined as 
\begin{equation}
    \tau_\textrm{acc} = \frac{t_\textrm{Edd}}{(1-\epsilon) \dot{m}}
\end{equation}
and $t_\textrm{Edd} \equiv M c^2/L_\textrm{Edd} \approx 4.5\times10^8$~years is the Eddington time. Accretion feeds the BH mass as well as angular momentum. With the accretion rate $\dot{M}_\textrm{acc} $, the  spin angular momentum of the BH evolves as~\citep{Bardeen:1970zz, sr1Brito:2014wla}
\begin{equation}\label{eqn_Jacc}
    \dot{J}_\textrm{acc} = \frac{j_\textrm{isco}}{e_\textrm{isco}} \dot{M}_\textrm{acc},
\end{equation}
where $j_\textrm{isco} = \frac{2M}{3\sqrt{3}}(1+2\sqrt{\frac{3r_\textrm{isco}}{M} -2}~)$ denotes the orbital angular momentum per unit mass of a test particle at ISCO. Thus the spin parameter $\tilde{a}$ can change with time due to accretion and can reach up to~$0.998$ in the case of an astrophysical black hole~\citep{maxatildeThorne}\footnote{This limit holds for thin disk models. For thick disk, the limit may change~\citep{maxatildeThickdisk}.}.

In reality, it is well known that an AGN evolves through multiple cycles of activity with an overall lifetime ranging from $10^7 - 10^9$~years~\citep{1982MNRAS.200..115S, Martini:2000mj, 2002MNRAS.335..965Y, 2004MNRAS.351..169M, Schawinski:2015xka, Turner:2015cma}. However, for simplicity of tracking the evolution of the SMBH, we consider $\dot{m}$ to be constant over the lifetime of an AGN and can be thought of as some average value. Hence, the black holes accreting uninterruptedly (constant $\dot{m}$) and evolving with time will grow exponentially as
\begin{equation}
    M = M_0 e^{\frac{t}{\tau_\textrm{acc}}}.
\end{equation}

Before going into the complete superradiance evolution, let us first gain an intuitive understanding of the relevant region on the parameter space $(\tilde{a},~M)$ such that a black hole with parameters lying in the region can spin down within the age of the universe. For a non-accreting SMBH ($\dot{m}=0$), we extract the region where the superradiance time scales of the growth of the scalar field in a given $nlm$ mode are shorter than the age of the universe using $\tau_{nlm}<T_\textrm{univ}\sim 10^{10}$~years. In fig.~\ref{fig:Depletion}, the three regions above the curves (solid lines) shown in three different panels correspond to the three most dominant modes ($nlm=011,~022,~033$) considered in this study. For a specific $\mu$, this gives the maximum possible parameter space wherein a black hole can potentially decrease its spin due to superradiance. It should be noted that the SR-active mass range given by tab.~\ref{table_1} is essentially the maximum mass range for the extremal spin shown in the figure.

For the growth of a scalar field around an accreting SMBH ($\dot{m}\neq0$), we further refine the parameter space $(\tilde{a},~M)$ by applying the criteria $\tau_{nlm}< \textrm{min}(\tau_\textrm{acc}, ~T_\textrm{univ})$ which yield a refined region above the curves drawn with the broken lines in fig.~\ref{fig:Depletion}. For a scalar mass $\mu$, such refined regions calculated for every three modes represent the maximum parameter space where the superradiant spin-down rate will dominate over the spin-up rate due to accretion. However, a black hole requires more time than $\tau_{nlm}$ to fully spin down to the critical spin, and hence the actual region depleted from black holes will be a subset of this region which will be discussed in the next section. 

This simple analysis indicates that the lower mass segment of the depletion region will be particularly affected by accretion physics, where the timescale of superradiance exceeds that of accretion. Moreover, if the central SMBH in the AGN does not fall within the abovementioned regions in fig.~\ref{fig:Depletion}, then a substantial accretion rate can drag the black hole into the designated region within the lifespan of the AGN, and eventually make the central SMBH an \sra SMBH. Thus the higher accretion rate in AGNs facilitates the generation of the scalar cloud and subsequent spin-down of the black holes which are not initially SR-active.

\subsection{Evolution of accreting supermassive black hole}
\label{subsec:evolutioneqns}
In order to investigate the accreting \sra SMBH, we solve simultaneous time-evolution equations derived from the conservation of energy and angular momentum of the system comprising the SMBH, the scalar field, and the accretion disk around the BH. The \sr timescale involved is much greater than the dynamical time scale of the BH, hence the evolution equations can be written in terms of instantaneous rates under quasi-adiabatic approximation and given as~\citep{sr00Brito:2015oca, sr26Ficarra:2018rfu},
\begin{subequations}\label{eq:EvolveWithAcc}
\begin{align}
 \frac{d M}{d t}&=-\sum_{nlm}2 M_s^{nlm}\omega_{I}^{nlm}+\dot{M}_\textrm{acc}~,\\
 \frac{d J}{d t}&=-\sum_{nlm}\frac{2}{\omega_{R}^{nlm}}m M_s^{nlm}\omega_{I}^{nlm}+\dot{J}_\textrm{acc}~, \\
 \frac{d M_s^{nlm}}{d t}&=2 M_s^{nlm}\omega_{I}^{nlm}-\dot{E}_\mrm{GW}^{nlm}~,\\
 \frac{d J_s^{nlm}}{d t}&=\frac{2}{\omega_{R}^{nlm}}m M_s^{nlm}\omega_{I}^{nlm}-\frac{1}{\omega_{R}^{nlm}}m\dot{E}_\mrm{GW}^{nlm}~.
\end{align}
\end{subequations}
Here $M$ and $J$ denote the instantaneous mass and spin angular momentum of the SMBH respectively. $M_s^{nlm}$ and $J_s^{nlm}$ represent the mass and orbital angular momentum of the scalar cloud for the superradiant mode $nlm$ at any instant. The rates of mass ($\dot{M}_\textrm{acc}$) and spin angular momentum ($\dot{J}_\textrm{acc}$) accumulation by the SMBH from the accretion disk are given by eq.~\ref{eqn_Macc} and eq.~\ref{eqn_Jacc} respectively. Moreover, $\omega_I^{nlm}$ and $\omega_R^{nlm}$ denote the SR growth rate and energy eigenvalue of $nlm$  mode, as given by eq.~\ref{eqn_srrate} and eq.~\ref{eqn_wr} respectively. We also account for the emission of gravitational waves from the scalar cloud, with the rate ($\dot{E}_\mrm{GW}^{nlm}$) provided by the eq.~\ref{eqn_Egwrate}. 

In our analysis, we assume the existence of a single scalar field in the universe, characterized solely by the mass parameter $\mu$. Although interactions of the scalar field could significantly alter the evolution~\citep{Yoshino:2012kn, Yoshino:2015nsa, Fukuda:2019ewf, sr4Baryakhtar:2020gao}, we simplify by considering no self-interaction or interaction with other standard model particles when investigating the influence of accretion on superradiant growth.

\begin{figure}
 \center
\includegraphics[width=\columnwidth]{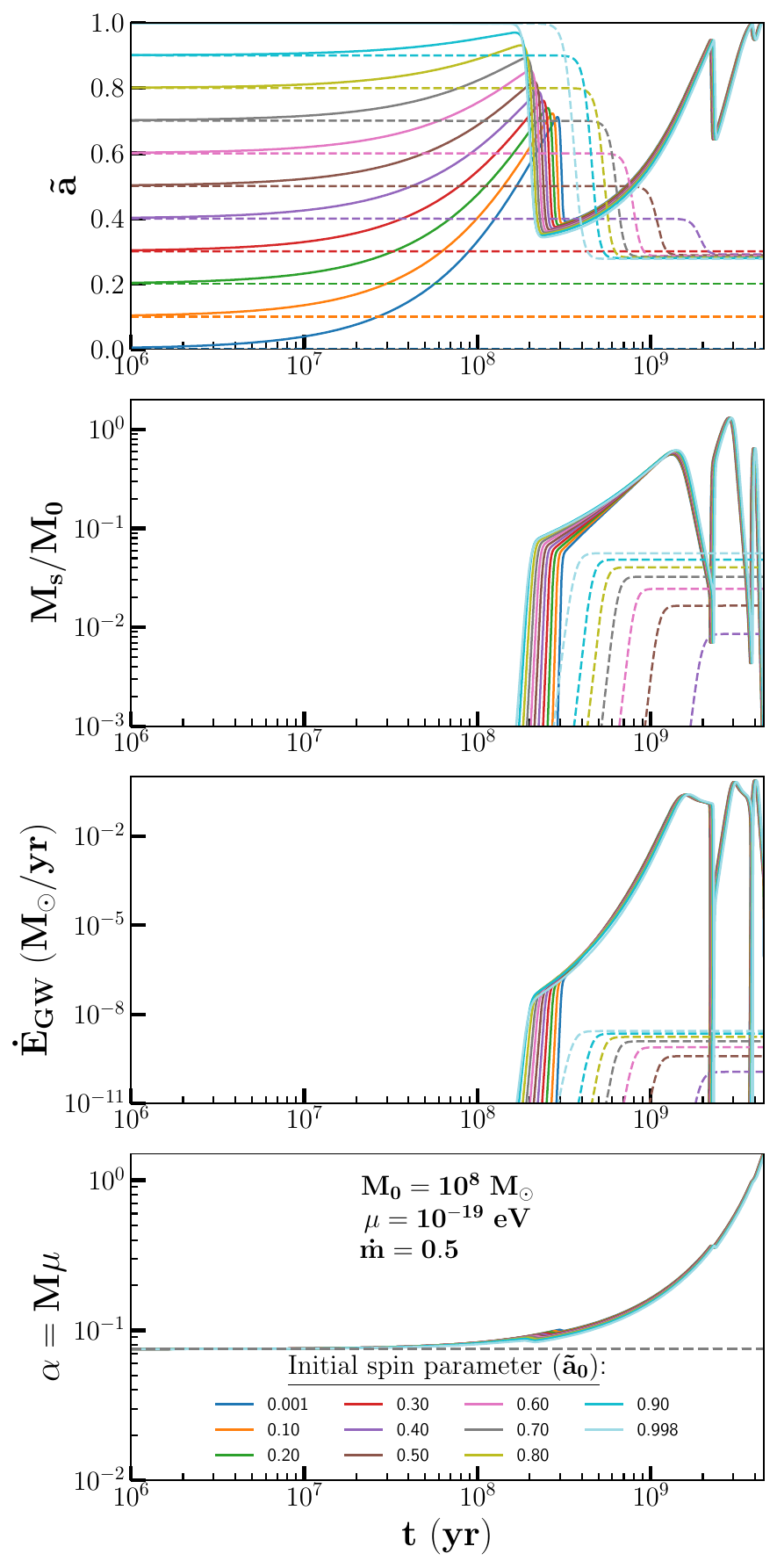}
  \caption{This figure shows an example of the time evolution of an accreting SMBH (solid lines) obtained by solving coupled differential eq.~\ref{eq:EvolveWithAcc}, in the presence of scalar field instability in the dominant modes $nlm=011,022,033$ with scalar mass $\mu=10^{-19}$~eV. The SMBH with different initial spin parameter $\tilde{a}_0$ and fixed initial mass $M_0 = 10^8~M_\odot$ at $t=0$ is assumed to accrete matter with accretion rate parameter $\dot{m}=0.5$. The time evolution of the spin parameter $\tilde{a}$, the total mass of the scalar cloud normalized to the initial BH mass $M_s/M_0$, gravitational wave emission rate $\dot{E}_{\mrm{GW}}$, and gravitational coupling $\alpha$ are shown in the four panels, arranged from top to bottom. The key effects of accretion on the superradiant evolution are- BH below the critical spin becomes SR-active by spinning up, an amplified growth of the scalar cloud, 8 orders of enhancement in $\dot{E}_{\mrm{GW}}$ as $\alpha$ increases, and preponing the appearance of all the three modes which are in contrast with a non-accreting case (dashed lines). A detailed discussion can be found in the text.}
  \label{fig:Evolution}
\end{figure}

We choose to demonstrate the evolution of the central SMBH within an AGN while the superradiant instability of the scalar field with mass $\mu=10^{-19}$~eV occurs in the vicinity of the SMBH. We consider an SMBH with initial mass $M_0=10^{8}~M_\odot$ lying in the SR-active mass range, alongside various possible initial spin $\tilde{a}_0$ spanning from $0$ to $0.998$ at $t=0$. Setting the accretion rate parameter $\dot{m}$ to 0.5 such that half the energy of the maximum possible luminosity ($L_\textrm{Edd}$) is accreted by the black hole. This corresponds to the Eddington ratio $f_\textrm{Edd}$ given by eq.~\ref{eq:fedd}, lying approximately in the range $0.03$ to $0.15$ depending upon the spin of the black hole. This range falls in the observational range $f_\textrm{Edd}\in (10^{-2}, 1)$ of AGNs as observed by SDSS~\citep{2020ApJS..249...17R}. Notice that it implies that the rate of mass accretion ($\dot{M}_\textrm{acc} = \dot{m}L_\textrm{Edd}/c^2$) increases as the BH mass increases due to the linear relationship between Eddington luminosity and BH mass. 

We begin our analysis at $t=0$ when the central SMBH in an AGN initiates its uninterrupted accretion phase (i.e. $\dot{m}$ remains constant over time). The total time available to evolve the SMBH must not only be less than the age of the Universe ($\approx 1.38\times10^{10}$~years according to the $\Lambda$CDM model, with $H_0$=67.66 km~s$^{-1}$~Mpc$^{-1}$, $\Omega_{0m}$= 0.31~\citep{Planck:2018vyg}) but also less than the time span following the formation of SMBHs. Although the exact formation epoch of the SMBH is still uncertain, the epoch of SMBH formation can be approximated to occur post the initial galaxy formation and between the redshifts 30 to 10. Thus, the cosmic time available for superradiant evolution could be approximately $10^{10}$~years. Conversely, as previously discussed, the typical lifespan of an AGN can extend to roughly $10^9$~years. Therefore, we limit the evolution of the equations to less than $10^{10}$ years. 

We only account for the growth of the three leading orders $(l=m=1,2,3)$ and the most dominant $(n=0)$ modes of the superradiant cloud. We consider the seed field originates from quantum fluctuations of the scalar field around the BH, thus setting the initial mass of the cloud for each mode to a negligible value ($10^{-10}~M_0$) compared to the SMBH, ensuring the final growth remains independent of the initial cloud mass\footnote{We refrain from delving into scenarios where the identical scalar particles act as dark matter, which would necessitate consideration of initial scalar abundance sourced from the galactic center's dark matter density and it may involve mode mixing if the initial abundance in a mode is significant~\citep{sr26Ficarra:2018rfu}. Such considerations lie beyond the scope of this study.}. We ignore the backreaction of the scalar cloud on the metric, for the reason that the scalar energy density is not large enough to change the structure of the geometry, as also argued in refs.~\citep{sr1Brito:2014wla, sr20Arvanitaki:2010sy, sr19Roy:2021uye}. 

In figure~\ref{fig:Evolution}, we present the numerical solution illustrating the time evolution described by eq.~\ref{eq:EvolveWithAcc} with various initial spin parameter ($\tilde{a}_0$) values of an SMBH and are all initialized with a mass of $10^8~M_\odot$ at $t=0$. These $\tilde{a}_0$ values are represented by different colors. This illustration provides insights into the typical time evolution of an accreting SMBH exhibiting superradiant activity induced by a scalar particle of mass $\mu=10^{-19}$~eV. Each panel displays solid curves representing superradiant evolution under the influence of accretion. Additionally, we include a comparison with a non-accreting scenario of superradiance, solved with the same initial conditions, represented by dashed curves in each panel.

The uppermost panel demonstrates the variation of $\tilde{a}$ over time. The second panel presents the time evolution of the total scalar cloud mass ($M_s = M_s^{011}+M_s^{022}+M_s^{033}$) relative to $M_0$. The third panel depicts the total gravitational wave emission rate ($\dot{E}_{GW} = \dot{E}_{GW}^{011}+\dot{E}_{GW}^{022}+\dot{E}_{GW}^{033}$) arising from the annihilation of scalar particles within the cloud into gravitons. Lastly, the bottom panel tracks the variation in the gravitational coupling parameter $\alpha = M\mu$ over time, primarily increasing due to mass accumulation by the SMBH from the accretion disk. However, occasional tiny drops occur in mass when the superradiant instability emerges successively in the three modes. The gray dashed line represents the initial value of $\alpha \approx 0.075$.

In the absence of accretion, only the 011-mode of the scalar cloud grows within the lifetime of an AGN. This is because the SR timescale is shortest for the 011-mode for the chosen $\alpha_0 = M_0\mu = 0.075$. As the field grows, it leaves the BH at $\tilde{a}_{\mrm{crit}}\sim 0.3$ and further growth stops in the 011-mode. Now, the SR timescale of the 022-mode and the 033-mode for this critical spin is in fact more than the age of the universe, which is why these higher modes do not grow significantly, and hence no further spin-down is observed. It can be seen in the figure that the SR time-scale for the 011-mode is minimum ($\sim 2\times 10^8$~years) for the extremal initial spin parameter 0.998 (light-cyan dashed-line), and superradiant growth for smaller $\tilde{a}_0$ dominates much later because the SR timescale is longer for them compared to 0.998. The maximum mass of the scalar cloud $M_s$ is largest for $\tilde{a}_0 = 0.998$ and remains less than 10\% of the BH mass whereas it is even smaller for $\tilde{a}_0$ lower than  0.998. During the evolution, the reduction of the scalar cloud caused by GW emission consistently remains subdominant, thus not impacting the growth of superradiance. We do not observe any spin-down or growth of the cloud for the initial spin parameters below the critical value $\approx0.3$ as they do not satisfy the SR condition given in eq.~\ref{eqn_srcondition}. 

For an accreting BH, all three modes of the scalar field can grow within the maximum possible lifetime ($\mathcal{O}(10^9)$ years) of an AGN which can be seen in fig.~\ref{fig:Evolution}. This is because $\alpha = M\mu$  increases with time as the BH accretes matter. Accretion drives the SR-active BH through various mass ranges as described in tab.~\ref{table_1} such that the timescale of the growth of the field in 011, 022, and 033 modes reduces drastically. It is important to note that when the initial spin is below the critical value given by eq.~\ref{eq:acrit}, the seed field generated by quantum fluctuations remains non-superradiant, leading to exponential suppression of its growth. Although the dynamics of the field is quite insensitive to this initial value, the challenge arises from the exponential decay of the cloud in the non-superradiant phase, which likely reduces the particle occupation number to less than one. This inhibits a classical treatment and necessitates the assumption that quantum fluctuations maintain the particle number near unity over sufficiently short time scales smaller than superradiant growth time scale. The panel with the plot of $M_s/M_0$ vs. $t$ shows the three peaks corresponding to the three modes 011, 022, and 033 growing consecutively as time increases. The three epochs of the beginning of these modes can be marked in the figure as the three dips in the spin versus time plot. The superradiant growth in any mode can be divided into five phases depending upon which physical process is dominant.

\begin{enumerate}
    \item \textit{Accretion-dominated phase:} This phase occurs when the accretion rate is larger than the superradiance rate. In fig.~\ref{fig:Evolution}, the initial mass of the BH is such that its initial evolution is dominated by accretion. Thus the BH first spins up irrespective of $\tilde{a}_0$ until superradiance takes over and the spin-down occurs as shown in the figure. The key role of accretion in this phase is spinning up the black hole with spin lying below the critical value and converting it to SR-active. 
    
    \item \textit{Superradiance-dominated phase:} During this phase the rate of superradiant growth becomes more than the accretion rate, and the GW emission from the cloud remains subdominant. The BH predominantly loses its spin angular momentum and an exponential growth of the scalar cloud occurs around the BH. It continues until the spin approaches the critical spin $a_\textrm{crit}$, but accretion keeps the spin just above the critical spin such that the evolution enters the next phase of an attractor phase.

    \item \textit{Attractor phase:} In this phase, the black hole's spin tends to converge at the critical value $a_\textrm{crit}$ due to the interplay between two competing processes of superradiant instability and accretion. This phase has previously been referred to as the over-threshold-superradiant phase in ref.~\citep{sr9Hui:2022sri}. As matter accretes onto the black hole, its spin increases, but the superradiant growth of scalar fields acts as a feedback mechanism, pulling the spin back toward the critical value. This dynamic interaction results in the black hole's spin persistently hovering slightly above $a_\textrm{crit}$ for an extended duration. Consequently, the scalar cloud continues to grow and exceeds 10\% of the black hole's mass, a threshold determined numerically as the upper limit for dominant mode growth around an isolated black hole~\citep{Herdeiro:2021znw}. However, the growth of the cloud in a given mode halts when the GW emission rate from the cloud becomes dominant, leading to the peak of $M_s/M_0$ as illustrated in fig.\ref{fig:Evolution}. 
    
    \item \textit{GW-dominated phase:} This phase begins after $M_s/M_0$ reaches its maximum in a given mode, where the cloud's growth rate matches its depletion rate due to GW emission. Notably, we observe an eight-order increase in the peak GW emission rate when accretion is present compared to an isolated black hole. This enhancement arises from the dependence of the GW rate on the mass fraction of the cloud and gravitational coupling, $\alpha$, given as $\dot{E}_\mrm{GW}\sim (M_s/M)^2 \alpha^{4l+10}$. During the 011-active phase of the SMBH, $\alpha$ increases from 0.075 to 0.25 when $M_s/M_0$ reaches its maximum, boosting the GW rate by seven orders. Additionally, there is another order of enhancement due to a 2.5 times increase in $M_s/M$ compared to without accretion. One can obtain the boosted observable strain $h$ at the GW detector near earth at a distance of $D$ from the AGN using~\citep{sr20Arvanitaki:2010sy, sr11Arvanitaki:2014wva, Brito:2017zvb, Ng:2020jqd} 
    \begin{equation}
        h = \sqrt{\frac{4 G c^2 \dot{E}_\textrm{GW}}{D^2 \omega^2}}.
    \end{equation}
    Therefore, the enhancement factor for the observable strain is approximately four orders of magnitude. This indicates that accretion-induced superradiance significantly amplifies the GW signature of superradiance, the extent of which relies on the strength of the accretion rate. Here, we showcase the enhancement in the GW rate for $\dot{m}=0.5$, corresponding to the Eddington ratio of an AGN ranging between 0.03 to 0.16, consistent with typically observed AGN Eddington ratios $f_\textrm{Edd}\in (10^{-2}, 1)$ by SDSS~\citep{2020ApJS..249...17R}. It is worth recalling that we consider $\dot{E}_{GW}$ in the flat background metric. For the case of the Schwarzschild metric, the GW emission rate is higher compared to the flat metric because of a larger prefactor \citep{sr00Brito:2015oca}. The percentage change in  $\dot{E}_{GW}$ could be as large as 100\% for the particular benchmark point shown here, implying that our prediction of the boost in $\dot{E}_{GW}$ due to accretion is underestimated in the flat metric assumption.
    
    \item \textit{Collapse phase:} This phase arises when the mass of the BH is increased so that the scalar cloud configuration, which was the eigenstate of the smaller mass BH previously, is no longer the eigenstate of the heavy mass BH. Therefore, the whole cloud in that mode shrinks and gives back the mass and angular momentum to the BH. However, by the time it collapses, GW emission has already reduced the cloud to about a percent of the BH mass, and hence, it is hard to see the change in the spin and mass of the BH in the plot. It could be interesting to see if this sudden collapse phase of the cloud could be a new source of GW bursts which can only occur if there is accretion-induced superradiance at the center of an AGN. However, it is beyond the scope of this study to further investigate this potentially new but subdominant GW signature of the SR cloud.
\end{enumerate}

In the example shown in fig.~\ref{fig:Evolution}, the BH will be SR-inactive after time $\sim 4\times10^9$~years when all three modes have grown up and collapsed and BH approaches the maximal spin. Hence, an \sra phase of the SMBH with initial mass $10^8~M_\odot$ with $\dot{m}=0.5$ would last for $\sim 4\times10^9$~years irrespective of its initial spin in the presence of a scalar field with mass $10^{-19}$~eV.

In summary, the SMBH experiences a decrease in spin during the superradiant growth of the scalar cloud, while accretion drives its spin upward. Once the first mode of the scalar cloud emerges, it leaves the black hole near the critical spin. Subsequently, the black hole enters an attractor phase followed by GW dominated phase where its mass and spin evolve, maintaining the spin slightly above the critical value while accreting until the next dominant mode emerges. Upon dominance of the next mode, the spin jumps to the critical value corresponding to that mode. Throughout the entire superradiant phase, the SMBH's mass increases by approximately an order of magnitude until $\alpha$ surpasses the upper limit of $3/2$ for the 033-active mass range, leading to an exponentially suppressed growth of the scalar field further~\citep{sr0Zouros:1979iw}. The influence of accretion-induced SR-active evolution of the SMBH can also affect the characteristics of the host AGN which we now discuss in the subsequent subsection.

\subsection{Characterstics of superradiant active AGN}\label{subsec:Lumandfedd}
Here we derive the characteristics of an AGN hosting an SR-active SMBH at the core which evolves with time. We mainly discuss the effects of spin-down on the Eddington ratio and luminosity in various wavelength bands. We use the Navikov-Thorne (NT) model of the accretion disk to calculate the luminosities in the X-ray (0.001-0.01~$\mu$m), UV (0.01-0.4~$\mu$m), and Vis-IR (0.4-100~$\mu$m) bands and see how they evolve with time.
\subsubsection{Eddington ratio}
Firstly we discuss the consequence of spin-down on the Eddington ratio $f_\textrm{Edd}$ of the SR-active AGN. To calculate the ratio, we estimate the bolometric luminosity $L$ using eq.~\ref{eq:Lbol} and the Eddington luminosity $L_\textrm{Edd}$ using eq.~\ref{eq:LEdd}.

\begin{figure}
    \centering
    \includegraphics[width=\columnwidth]{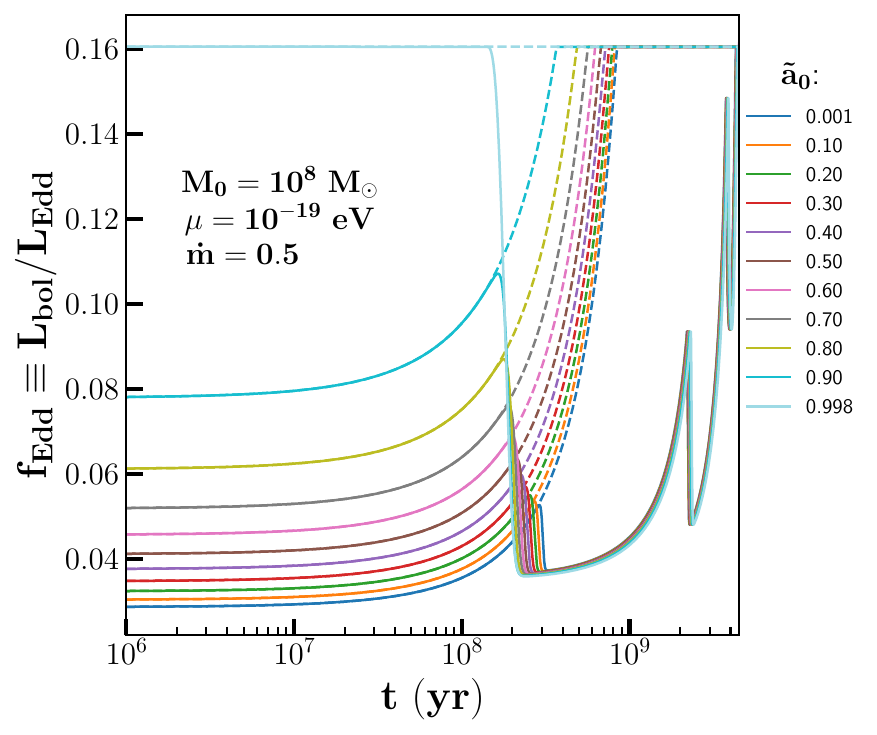}
    \caption{This figure shows the evolution of the Eddington ratio $f_\textrm{Edd}$ (solid lines) of an \sra AGN of various initial spins shown by different colors. For comparison, we also show how $f_\textrm{Edd}$ will behave without \sr $i.e.$ in the presence of accretion only(dashed lines). Without \sr $f_\textrm{Edd}$ tends to grow monotonically, whereas it experiences sudden drops due to superradiance.}
    \label{fig:EddingtonRatio}
\end{figure}

Figure~\ref{fig:EddingtonRatio} shows the evolution of the Eddington ratio for various initial spins $\ta_0$ (solid lines) calculated using the instantaneous mass and the spin of the SMBH obtained from the representative example of accretion-induced superradiance as illustrated in fig.~\ref{fig:Evolution}. We observe sudden drops in the Eddington ratio of the AGNs at the time-scales corresponding to various modes of superradiant growth. This behavior of the evolution can be understood from the fact that the Eddington ratio is proportional to the radiative efficiency $\epsilon(\ta)$, which experiences these sudden falls due to the spin-down. Another outcome of the drop in the radiative efficiency can be seen in the growth of the SMBH. The growth rate of the SMBH, as can be seen from the eq.~\ref{eqn_Macc}, is proportional to $1-\epsilon$. Therefore, every time $\epsilon(\ta)$ experiences a sudden drop due to the spin-down, there will be a corresponding boost in the SMBH growth.

For comparison, we also show how $f_\textrm{Edd}$ will evolve with accretion only, depicted by the dashed lines. One can see that in the absence of the scalar field in our Universe, the Eddington ratio for $\ta_0<0.998$ monotonically increases with time due to accretion and asymptotically reaches a value $f_\textrm{Edd, max}\sim 0.16$ for $\dot{m}=0.5$ as the BH approaches to the near-extremal spin of 0.998. For the near-extremal spin, the evolution is flat because the spin does not change anymore once it reaches the maximal value, and hence the radiative efficiency that determines  $f_\textrm{Edd}$ for the constant accretion rate parameter $\dot{m}$ will remain flat throughout the evolution. In the absence of the scalar, it can be seen that after about $10^8$~years all the AGNs would be found to have Eddington ratios $f_\textrm{Edd, max}$ corresponding to the maximal spin. In contrast to this, the evolution will no longer be monotonically increasing in the presence of \srI due to the scalar, rather, it will fall (due to superradiance) and rise (due to accretion) at various epochs. For the particular benchmark value of the scalar mass, we can see that $f_\textrm{Edd}$ of the SR-active AGN reaching the maximal value will be delayed to $\sim 4\times10^9$~years. It should also be noted that the growth of the SMBH in the presence of accretion with superradiance is larger than the growth with accretion only.

\subsubsection{Luminosities in X-ray, UV, and Vis-IR bands}
\begin{figure*}
 \center
 \includegraphics[width=\columnwidth]{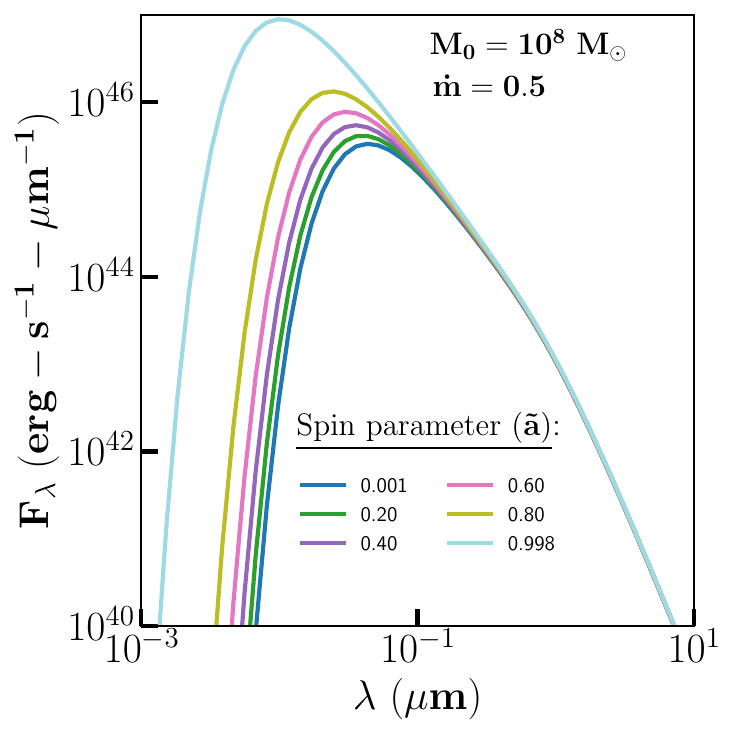}
  \includegraphics[width=\columnwidth]{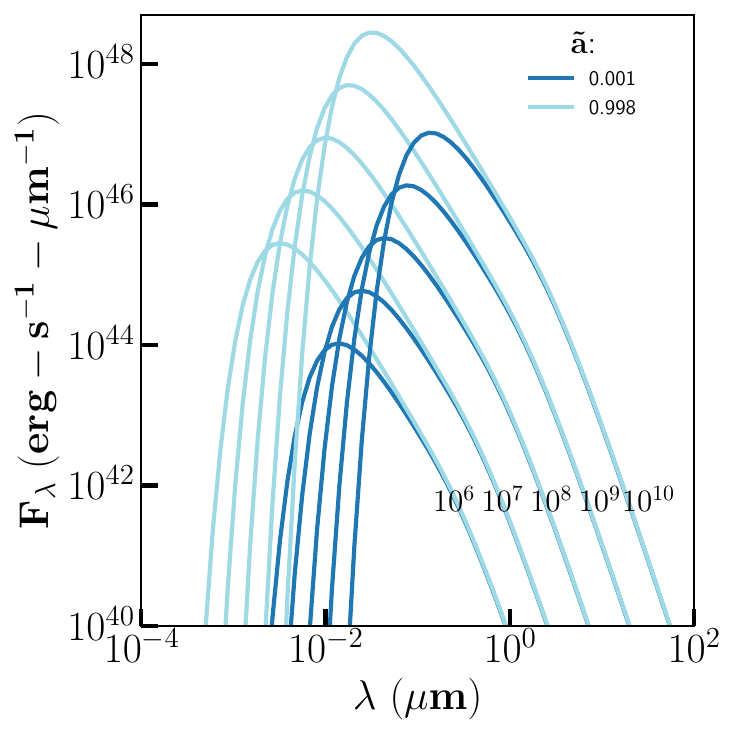}
  \caption{This figure depicts the spin and mass dependence of the color-corrected spectrum of the radiation from an  AGN disk in its rest frame. The dimensionless accretion rate parameter $\dot{m}$ and and the spectral hardening factor $f_{\mrm{col}}$ are assumed to be $0.5$  and $1.7$ respectively. In the \textit{left panel}, we show how the spectrum changes with the spin of the \sra SMBH with mass $10^{8}~M_\odot$ at the center of AGN. The effect of spin-down is most visible in the wavelength range less than $0.4~\mu$m which lies in X-ray and UV bands. In the \textit{right panel}, we show the spectrum for different values of the mass of the SMBH annotated on the curves. The two different shades of the color light blue and dark blue represent two benchmark values of spins 0.001 and 0.998 respectively. This illustrates how superradiant instability around SMBHs can affect the radiation emitted within the wavelength range of $10^{-4}~\mu$m to 100$~\mu$m from the thin disk in the AGNs.}
  \label{fig:Lumperunitwavelength}
\end{figure*}

The physics of the steady-state accretion disk around a spinning black hole is best explained by the NT model. This model falls into the category of accretion disk model termed as \ita{thin disk model} where it is assumed that the thickness of the disk is always less than the radial size. The thin disk model is valid as long as the accretion rate parameter $\dot{m}\in (0.01,10)$~\citep{Abramowicz:2011xu}. One of the key impacts of spin-down on the disk is increasing the radius of the inner edge of the disk which is expected to be at the innermost stable circular orbit (ISCO) around the black hole~\citep{sr00Brito:2015oca}. Additionally, it is only the inner part of the disk (up to $\sim 10$ gravitational radii) that is highly influenced by the spin down. Moreover, for SMBHs, the typical time scale of the superradiant spin-down is greater than 100 years, which is much larger than the orbital period ($\sim~1$~day) of test particles near the ISCO. We, therefore, assume that the superradiant spin-down will keep the accretion disk in a quasi-static equilibrium and hence make use of the  NT model of the accretion disk to calculate the time variation in the AGN characteristics.

The NT model yields the flux emitted from the disk's surface as a function of the radial distance and spin parameter $\ta$ of the BH. In the Kerr background, the expression for the flux goes as~\citep{Abramowicz:2011xu}
\begin{equation}\label{eqn_FluxNT}
F (r) = 7 \times 10^{26} \frac{\mathrm{erg}}{\mathrm{s}\mathrm{~cm}^{2}} \dot{m} \frac{M_\odot}{M} \left(\frac{M}{r}\right)^3 \mathcal{B}^{-1} \mathcal{C}^{-1 / 2} \mathcal{Q}.
\end{equation}
where the radial distance $r$ is taken from the black hole center. There are mainly three parameters that go into the flux: the mass of the black hole $M$, the accretion rate parameter $\dot{m}$, and the spin parameter $\tilde{a}$. The radial functions $\mathcal{B}, \mathcal{C}, \mathcal{Q}$  are defined in terms of $y=$ $(r / M)^{1 / 2}$ and $\ta$ as:
\begin{align*}
\mathcal{B}&=1+\ta y^{-3}, \quad
\mathcal{C}=1-3 y^2+2 \ta y^{-3}, \\
\mathcal{Q}_0&=\frac{1+\ta y^{-3}}{y\left(1-3 y^{-2}+2 \ta y^{-3}\right)^{1 / 2}}, \quad
\mathcal{Q}= \mathcal{Q}_0\left(\mathcal{Q}_1 - \mathcal{Q}_2\right),\\
\mathcal{Q}_1&=y-y_0-\frac{3}{2} \ta \ln \left(\frac{y}{y_0}\right)-\frac{3\left(y_1-\ta\right)^2}{y_1\left(y_1-y_2\right)\left(y_1-y_3\right)} \ln \left(\frac{y-y_1}{y_0-y_1}\right),\\
\mathcal{Q}_2 &=\frac{3\left(y_2-\ta\right)^2}{y_2\left(y_2-y_1\right)\left(y_2-y_3\right)} \ln \left(\frac{y-y_2}{y_0-y_2}\right)
\\
&+\frac{3\left(y_3-\ta\right)^2}{y_3\left(y_3-y_1\right)\left(y_3-y_2\right)} \ln \left(\frac{y-y_3}{y_0-y_3}\right).
\end{align*}
Here $y_0=\sqrt{r_{\mathrm{isco}} / M}$, and $y_1, y_2$, and $y_3$ are the three roots of $y^3-3 y+2 \ta=0$; that is

\begin{align}
 y_1&=2 \cos \left[\left(\cos ^{-1} \ta-\pi\right) / 3\right], \nonumber\\
y_2&=2 \cos \left[\left(\cos ^{-1} \ta+\pi\right) / 3\right], \nonumber \\
y_3&=-2 \cos \left[\left(\cos ^{-1} \ta\right) / 3\right] \nonumber.    
\end{align}

Once the flux is obtained, one can find the spectrum from the disk by simply assuming that each point at $r$ on the disk behaves locally as a black body (BB) characterized by the local surface temperature $T_\textrm{s}(r)$. The local surface temperature can be estimated using the flux $F(r)$ coming from the surface given by eq.~\ref{eqn_FluxNT} and applying the Stefan-Boltzmann law, given as
\begin{align}
    T_\mathrm{s}(r)=[\sigma F(r)]^{\frac{1}{4}}.
    \end{align}
where $\sigma$ is the Stefan-Boltzmann constant. It should be noted that the electron scattering modifies the BB spectrum, specifically the radiation coming from the inner part of the accretion disk. Hence it is necessary to take into account this modification. One of the most popular ways of dealing with this tweak in the spectrum is through \ita{color correction} or \ita{spectral hardening} factor $f_{col}$. $f_\mathrm{col}$ being weakly sensitive to the mass of the black hole, can lie in the range $f_\mathrm{col}\sim 1.4-1.7$~\citep{1995ApJ...445..780S, Davis:2004jf, Davis:2006bk}. In our work, for simplicity, we use a fiducial value of $f_\mathrm{col}=1.7$. Hence the modified or the color-corrected  blackbody spectra can be given by~\citep{2011MNRAS.414.1183K, Davis:2004jf, Heydari_Fard_2023, Daly:2019srb}

\begin{equation}\label{eqn_modifiedflux}
    f_\lambda(r)=\frac{\pi}{f_\mathrm{col}^4}B_\lambda (f_\mathrm{col}T_\mathrm{s}(r)),
\end{equation}
where $B_\lambda$ is the Planck function for blackbody radiation given by, 
\begin{equation}
    B_\lambda(r) = \frac{2 h c^2}{\lambda^5}\frac{1}{e^\frac{h c}{\lambda k_B T_\mathrm{s}(r)}-1}.
\end{equation}

The AGN spectrum now can be calculated by integrating the modified BB spectrum (given by eq.~\ref{eqn_modifiedflux}) over the entire disk as
\begin{align}\label{eqn_lum}
  F_\lambda = 2\int f_\lambda(r) r drd\phi=4\pi\int f_\lambda(r) r dr.
\end{align}
Factor 2 in the above equation is considered to incorporate the flux coming from both sides of the disk. The radial integration is performed from $r_\textrm{isco}$ to $10^3 M$ within which the flux contribution is most dominant.

\begin{figure}
\centering  
\includegraphics[width=\columnwidth]{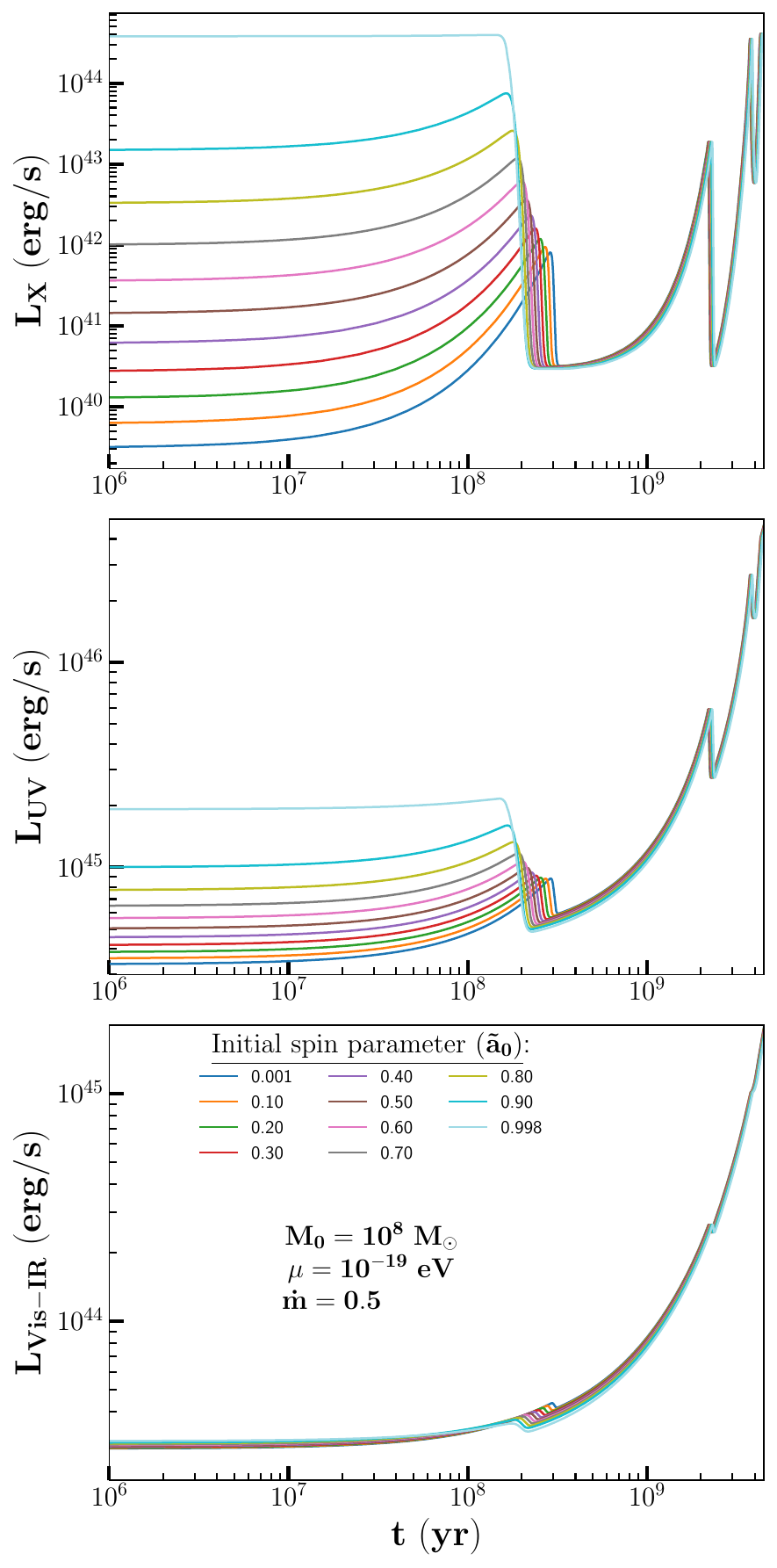}
\caption{This figure shows the derived evolution of the luminosities of the SR-active AGN in the X-ray ($10^{-4}-10^{-2}~\mu$m), UV (0.010-0.4~$\mu$m), and Vis-IR (0.4-100~$\mu$m) bands. We took the same example of the initial spin and mass as shown in fig.~\ref{fig:Evolution} for demonstrating the impact of superradiance on the time evolution of the AGN luminosities in various bands. In the \textit{top panel}, we plot the time-variation in the X-ray luminosity as the SR-active AGN evolves. As it is the higher energy band for which the impact of spin is highest (see fig.~\ref{fig:Lumperunitwavelength}), the change in the X-ray luminosity is maximum as compared to the UV and Vis-IR as the spin down occurs due to superradiance. In the \textit{middle panel}, we show the variation of luminosity in UV band where the AGNs are most luminous. The \textit{bottom panel} shows the time evolution of the luminosity in Vis-IR band where the impact of spin is least.}
  \label{fig:Lbands}
\end{figure}

In fig.~\ref{fig:Lumperunitwavelength}, we show the spin and mass dependence of the color-corrected spectrum of the radiation from an AGN in its rest frame. For the variation in the spin, we take an SMBH at the center of an AGN of mass  $10^{8}~M_\odot$ as a reference point. During spin-down, the spectrum squeezes towards the higher wavelength signifying the fact that there is a prominent depletion of the high-energy photons during the spin-down process. The effect could be understood intuitively from the fact that the high energy part of the spectrum dominantly comes from the ionized inner region of the disk where the impact of spin is highest. This implies that the higher the energy of the photon the larger its possibility to come from the location nearer to the black hole $i.e.$ near the ISCO. Furthermore, as the the spin of the black hole goes down to zero from the maximal value of $0.998$, the radius of ISCO also changes approximately from $ M$ to $6M$, and hence effectively the inner material shifts away from the black hole resulting in the reduction of the high energy photons. However, the impact of spin is minimal for the middle and outer region of the accretion disk from where the lower energy photons are being emitted. As a result of this,  the large wavelength part of the spectrum is least sensitive to the change in the spin and therefore dominantly determined by the mass of the SMBH. The effect of mass on the spectrum can be seen in the right panel of the figure. Here we observe a shift towards the higher wavelength as the black hole mass increases. Increasing black hole mass also increases the height of the peak of the spectrum which further implies that the AGN will become brighter as the black hole mass increases. 

Although the spectrum coming from the whole accretion disk turns out to be a continuum, the right panel tells us about the dominant part of the spectrum which lies between $10^{-4}~\mu\textrm{m}$ to 100~$\mu\textrm{m}$ for SMBHs at the core. One can therefore divide the full spectrum into three distinct wavelength bands which will be most sensitive to the spin-down of the SMBH in an SR-active AGN. We calculate the luminosities in the bands of- (1) X-ray ($10^{-4}-0.01~\mu$m), (2) UV  ($0.01-0.4~\mu$m), and (3) Vis-IR  ($0.4-100~\mu$m) by integrating the spectrum described in fig.~\ref{fig:Lumperunitwavelength} in a specific wavelength range relevant to the bands considered as follows,
\begin{subequations}\label{eq:Lbands}
\begin{align}
L_{\mrm{X}} =&  \int_{10^{-4}}^{0.01} F_\lambda d\lambda, \\
L_{\mrm{UV}} =&  \int_{0.01}^{0.4} F_\lambda d\lambda, \\
L_{\mrm{Vis-IR}} =&  \int_{0.4}^{100} F_\lambda d\lambda,
\end{align}
\end{subequations}
where the integration limits are in the unit of micrometer ($\mu$m). 

In fig.~\ref{fig:Lbands}, we show the results of the time evolution of the luminosities of the SR-active AGN in all three bands which are obtained using eqs.~\ref{eq:Lbands} for the representative example of the mass and spin evolution shown in fig.~\ref{fig:Evolution}. From the fig.~\ref{fig:Lbands}, one can notice that the \sra AGN will be most luminous in the UV band ($10^{44-46}$~erg/s), however as we can see the impact of spin down is maximally revealed in the  X-ray band. This is in agreement with our understanding that the higher energetic photons dominantly come from the innermost part of the accretion disk and hence highly affected by the spin-down. The drop in the luminosity in each band first occurs around $10^8$~years where the very first mode of the superradiant cloud originates at the expense of the spin angular momentum of the SMBH. The X-ray luminosity, in the first mode, alters by five orders in magnitude, whereas the change in the UV band luminosity is always below an order in magnitude. The value of X-ray luminosity after the first mode remains below $10^{41}$~erg/s until accretion again comes into play resulting in a rise in the mass and spin of the black hole. The luminosity then drops down at around $t\sim 10^9$~years due to the spin-down of the black hole or the growth of the cloud in the  $022$-mode. Consequently, the X-ray luminosity changes by two orders in magnitude and the change in the UV luminosity is again not more than an order. Following this, there is again a rise due to accretion and then a drop caused by the dominance of the $033$-mode. Although the luminosity of the SR-active AGN in the Vis-IR band is appreciable compared to the X-ray, it gives the least amount of information about the spin-down effect. Therefore, one can draw interesting effects of superradiance in the X-ray and UV band luminosities of AGNs.

\section{Distribution of superradiant active AGNs}
\label{sec:observables}
We now derive the effect of superradiance on the distribution of the characteristics of AGNs hosting SR-active SMBHs at the core. We first calculate the time evolution of the distribution of accreting SMBHs lying in the \sra mass range defined in table~\ref{table_1}. We consider $10^4$ samples of SMBHs whose logarithm of initial masses ($Log_{10}(M_0)$) and initial spins ($\tilde{a}_0$) are distributed uniformly at $t=0$. For each scalar mass $\mu$, we focus the analysis on the corresponding SR-active mass range of the black hole as described in tab.~\ref{table_1} and is such that the minimum mass corresponds to the mass for which 011-mode could be active and maximum mass for which 033-mode could be active within $T_\textrm{univ}$  \textit{i.e.} 
\begin{align*}
  M_0 \in [4.2 \times 10^{7} \msun \left(\frac{10^{-19}~\textrm{eV}}{\mu}\right)^{1/9},~2.0 \times10^{9} \msun \frac{10^{-19} \mrm{eV}}{\mu}].
\end{align*}
The initial spins of these BHs taken to be distributed uniformly between a very small value of $0.001$ and the maximum possible spin of an astrophysical black hole $0.998$.

We evolve these $10^{4}$ SMBHs sampled in the above-mentioned parameter space by numerically solving the time-evolution eqs.~\ref{eq:EvolveWithAcc}. We again consider the growth of superradiant cloud in three leading order modes ($nlm=011,022,033$) and depletion of the cloud through GW emission. We continue to take a constant accretion rate parameter to be at the benchmark value $\dot{m}=0.5$ for which the Eddington ratio will lie between 0.03 to 0.16 depending upon the spin of the SMBH (see fig.~\ref{fig:EddingtonRatio}). 

\subsection{Regge plane}\label{subsec:reggeplane}
\begin{figure*}
 \center
\includegraphics[width=\textwidth]{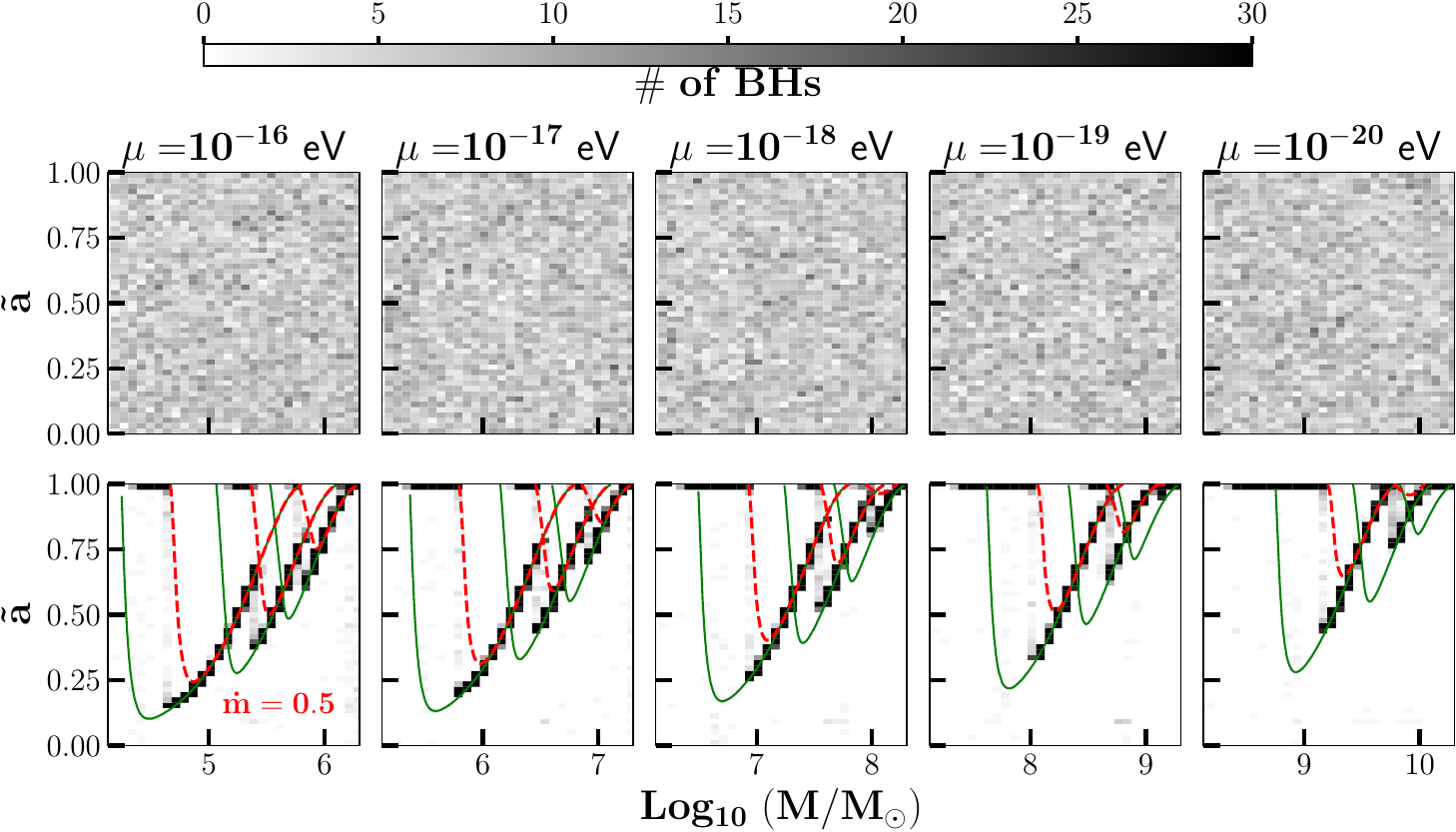}
  \caption{This figure depicts the appearance of the depletion region due to the superradiant evolution of the distribution of the accreting black holes on the spin-mass (Regge) plane. The accretion rate parameter $\dot{m}$ is taken to be 0.5. The evolution is shown for different scalar masses $\mu$ along different columns. The initial uniform distribution of $N=10^4$ SMBHs at $t=0$ superradiantly evolves to finally give the depletion regions in the Regge plane and accumulation of the BHs along the different tracks corresponding to $011,022,033$ modes. The fate of the whole population is decided by the time-scales- $\tau_\textrm{growth}^{nlm}$ and $\tau_\textrm{acc}$. BHs above the red-dashed line ($\tau_\textrm{acc}=\tau_\textrm{growth}^{nlm}$) for which $\tau_\textrm{growth}^{nlm}<\tau_\textrm{acc}$, first undergo \sr and then move along the $\tilde{a}_{\mrm{crit}}$ curve due to accretion, whereas BHs below the red line with $\tau_\textrm{acc}<\tau_\textrm{growth}^{nlm}$ will be initially dragged into the red-line due to accretion and then will experience similar superradiance. The green line shows the case for the non-accreting case with $\tau_\textrm{growth}^{nlm}=\tau_\textrm{univ}$. The difference between red (accreting) and green line (non-accreting) trajectories is revealed mostly for lower mass BHs for which the  \sr rate is smaller(see fig.~\ref{fig:Tau_sr}) and hence it is accretion that drives their fate.}
  \label{fig:Regge_hist}
\end{figure*}

In fig.~\ref{fig:Regge_hist}, we show the initial and final distributions of the \bhs in the Regge plane (\citep{sr20Arvanitaki:2010sy}) \ita{i.e.} spin vs. mass plane of the black hole. Each column in the figure corresponds to different scalar masses chosen to be $\mu \in \{10^{-16},10^{-17},10^{-18},  10^{-19},10^{-20}\}$~eV for illustration. As the distribution evolves, we observe the formation of gaps/depletion regions as the \sra \bhs start spinning down due to the growth of the scalar cloud in each of the modes. The black holes spin down to the boundary of the depletion region resulting in the accumulation along a curve approximated by the critical spin $\tilde{a}_\mrm{crit}(M,\mu, m)$ given by eq.~\ref{eq:acrit}. Notice, that the accumulation will not be exactly at $\tilde{a}_\mrm{crit}$ because of accretion as discussed in the previous section rather the spins will hover slightly above the critical value. 

For each mode, there are mainly two time-scales that decide the fate of this distribution - (1) the accretion time ($\tau_\textrm{acc}$) and (2) the time $\tau_\textrm{growth}^{nlm}$ required to grow the scalar cloud significantly. The growth time can be well estimated by taking the typical order of the mass of the scalar cloud, say 10\% of the black hole's mass and is given by
\begin{eqnarray}\label{eqn_srtimescale}
    \tau_\textrm{growth}^{nlm} \equiv \frac{\ln N_{\mrm{max}}}{2\omega_I^{nlm}} = \frac{1}{2\omega_I^{nlm}}\ln{\frac{0.1M c^2}{\mu}}.
\end{eqnarray}
Thus, black holes positioned in the region of the Regge plane where $\tau_\textrm{growth}^{nlm}<\tau_\textrm{acc}$ will initially experience spin reduction due to superradiant growth, settling their spin near $\tilde{a}_\mrm{crit}(M,\mu, m)$, and subsequently hovering it slightly above the critical value due to accretion. This population of black holes is marked by drawing a curve (red-dashed) defined as  $\tau_\textrm{growth}^{nlm}=\tau_\textrm{acc}$ in the Regge plane above which the superradiant spin down dominant over the spinning up due to accretion. Therefore, it is these black holes lying above the red-dashed curve that undergo superradiant spin down first and then spin up approximately along the critical spin ($\tilde{a}_{\mrm{crit}}$) curve due to accretion. It should be noted that the BHs can stay inside the red curve only until their lifetime $t=\tau_\textrm{growth}^{nlm}$. 
 
The population below the red-dashed curve pre-dominantly evolves due to accretion as for them  $\tau_\textrm{acc}<\tau_\textrm{growth}^{nlm}$. Because of accretion, these black holes get dragged above the critical spin and once they cross the red-dashed curve, they too undergo superradiant spin down within the time-scale of $\tau_\textrm{growth}^{nlm}$. This is why the left (lower mass) edge of the red-dashed curve of a given mode provides the lowest mass of accreting black holes which can be depleted and cluster along the critical spin $\tilde{a}_{\mrm{crit}}$. 

For analytical comparison with the non-accreting case ($\dot{m}=0$), we have also drawn a (green-solid) curve showing the previously obtained relevant parameter space defined by $\tau_\textrm{SR}^{nlm}=T_\textrm{univ}$ (as also shown in fig.~\ref{fig:Depletion}). All the BHs under this curve will undergo superradiant spin-down within $T_\textrm{univ}$. Thus, as discussed in \ref{subsec:accretionmodel}, we see from our numerical simulation that the depletion region caused by the accreting BHs is a subset of that is caused by the non-accreting ones. Here also one can make note of the fact that in each mode $nlm$, it is the lower end of the BH mass range that gets remarkably affected by accretion. This follows from the fact that for lower mass BH, $\tau_\textrm{growth}^{nlm}$ is larger as can be seen from fig.~\ref{fig:Tau_sr}. Hence these BH will essentially evolve due to accretion as $\tau_\textrm{acc}<\tau_\textrm{growth}^{nlm}$.

For each scalar mass, there exists a specific BH mass range corresponding to each mode as seen in table \ref{table_1}.  For the $011$ mode mass range, the accumulation along the boundary starts when the black holes enter into the $011$ mode of instability \ita{i.e.} after $t\approx \tau_{011}$.  The \bhs then start moving along the boundary due to accretion until they spin down to the critical spin of the $022$ mode, followed by which they now move along the boundary of the $022$ mode. This continues again until the scalar cloud grows in the $033$ mode, and eventually reaches the maximum value of spin $0.998$ because of accretion. For the $022$ mode mass range, the \bhs start spin-down in the $022$ mode after which they follow a similar trajectory as described for the $011$ mode. This explanation holds for the $033$ mode mass range as well. One can notice that the depletion region gets shrunk as the scalar mass decreases, this is intuitive from the fact that \sr growth rate is sufficient only for the BHs with higher spin and mass. 

One interesting feature of the superradiant evolution is that once the SMBHs spin down onto the boundary of the depletion region, there exists a spin-mass relationship defined by the $\tilde{a}_\textrm{crit}$ curve on the Regge plane. It can also be thought of as a \textit{synchronization} of spins of similar mass SMBHs. Hence, observation of synchronized spins of similar mass SMBHs may also hint towards its superradiant history.

\subsection{Luminosities in various bands}{\label{subsec:luminvariousbands}}
 
\begin{figure*}
\centering  
\includegraphics[width=\textwidth]{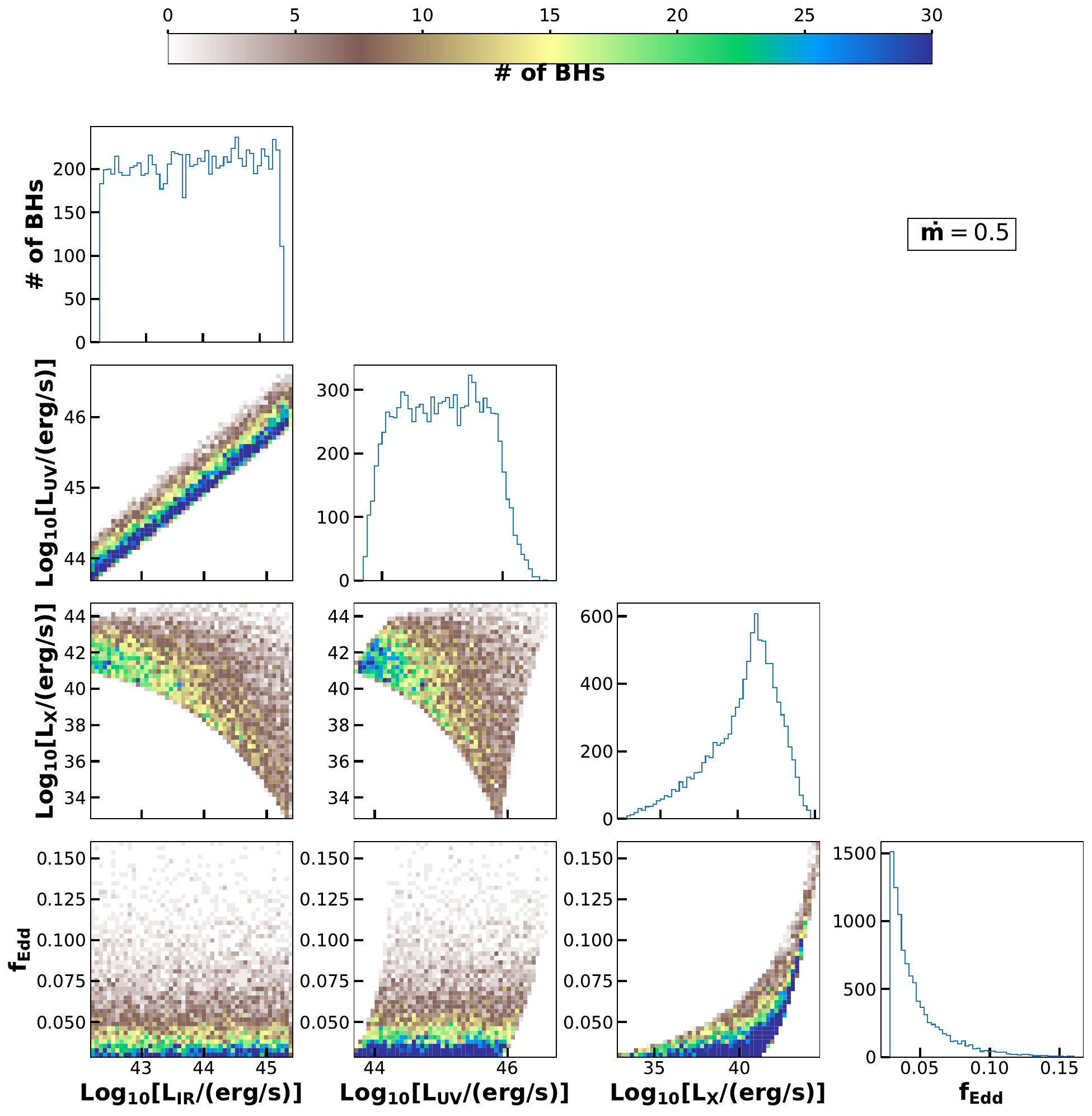}
\caption{This is the distribution of AGNs luminosities and Eddington ratio at $t=0$ with the dimensionless accretion rate parameter $\dot{m}=0.5$. This corresponds to the initial distribution of the SMBHs as shown in fig.~\ref{fig:Regge_hist}.}
  \label{fig:InitialLfeddbands1e-19}
\end{figure*}

\begin{figure*}
\centering  
\includegraphics[width=\textwidth]{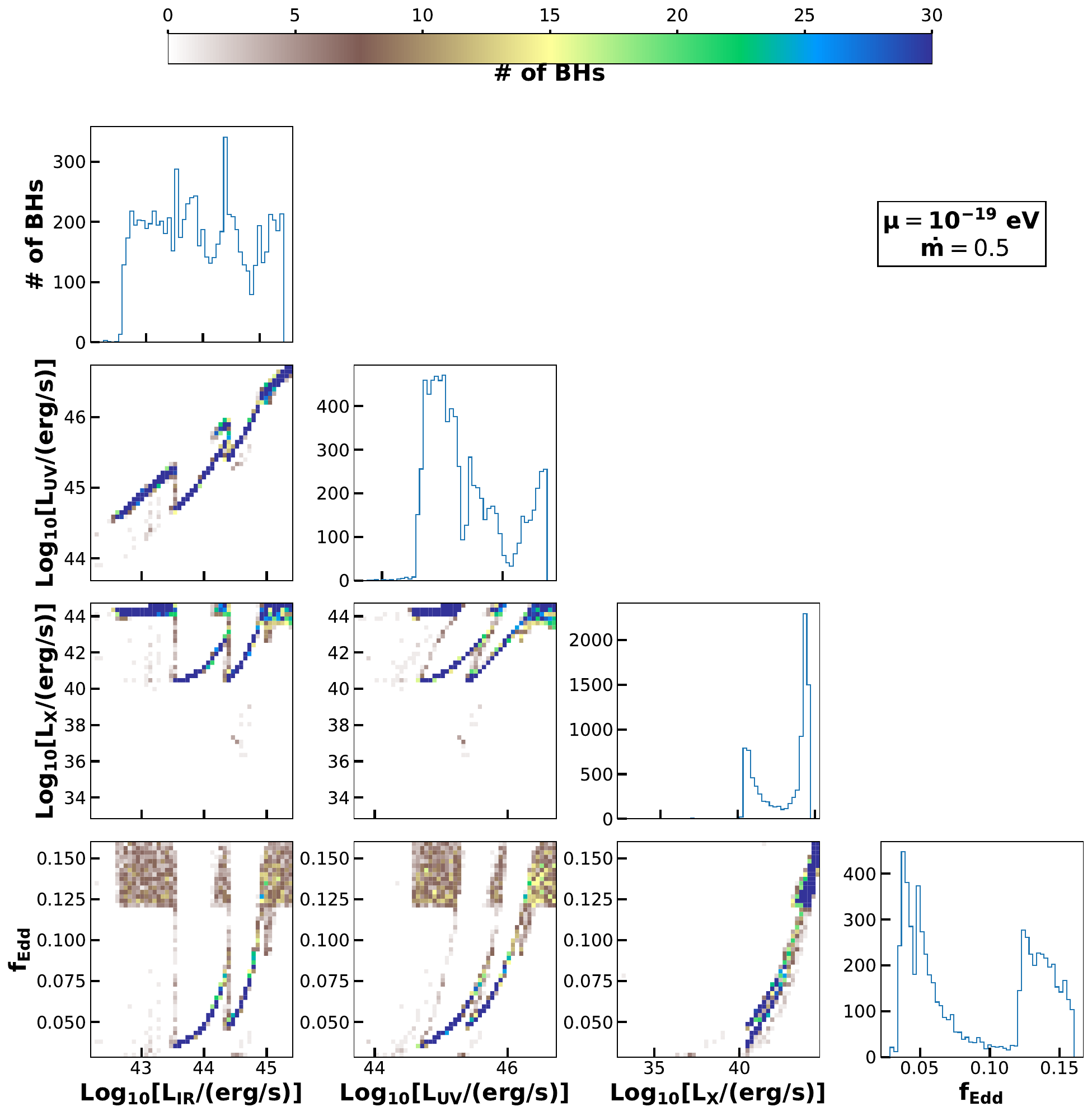}
\caption{ This is the distribution of AGN luminosities after $7\times10^8$~years ($\approx t_\textrm{Edd}$) of uninterrupted accretion phase with accretion rate parameter $\dot{m}=0.5$ in the presence of a scalar field in the universe with $\mu = 10^{-19}$~eV. Here one also can notice the formation of the depletion region and accumulation along various tracks corresponding to different modes, similar to fig.~\ref{fig:Regge_hist}. The depletion region is most visible in the $L_{\mrm{X}}-L_{\mrm{IR}}$, $ L_{\mrm{X}}-L_{\mrm{UV}}$ and  in the plane of $f_\textrm{Edd}$ vs different band-luminosities }
  \label{fig:Lfeddbands1e-19}
\end{figure*}
Hitherto we have noticed that the presence of the scalar field in our Universe forces the \sra SMBHs to follow the spin-mass relationship defined by the $\tilde{a}_\textrm{crit}$ (eq.~\ref{eq:acrit}) curve on the Regge plane. One can hence anticipate some kind of specific relation among the various  AGN band-luminosities from the dependence of the accretion disk on the spin and mass of the SMBH, showing the imprints of their \sr history.  Furthermore, the depletion region in the Regge plane should produce a depletion region on the plane of AGN luminosities in various bands. We use the NT model of the accretion disk to obtain the distribution of the luminosities of the AGNs with an \sra SMBH at the core. We follow the procedure described in subsec.~\ref{subsec:Lumandfedd} to calculate the Eddington ratio and the luminosities in X-ray, UV, and Vis-IR bands due to the continuum radiation from the accretion disk. 

In fig.~\ref{fig:Lfeddbands1e-19}, we show the distribution of the calculated Eddington ratio and the luminosities in X-ray, UV, and Vis-IR bands forming various activity planes of the AGNs. We perform the calculation for $10^4$ SMBHs uniformly distributed in the \sra mass range corresponding to the scalar mass in the range $10^{-16}-10^{-19}$~eV, but for consistency, we show the case for $\mu=10^{-19}$~eV. Their initial spins are also distributed uniformly in the range $[0.001,0.998]$. We choose to present the final distributions at $t=7\times 10^8$ years ($\approx t_\textrm{Edd}$), when we expect most of the \sra SMBHs to start following the spin-mass relationship along the boundary of the depletion region, to see if there exists any correlation among the quantities. 

At the time $t=0$, although the spins and masses are uniformly distributed  (see fig.~\ref{fig:Regge_hist}), their corresponding distributions of luminosities in various bands and the Eddington ratio are non-uniform as seen in fig.~\ref{fig:InitialLfeddbands1e-19}. This is because the variation in the mass and spin non-uniformly affects various parts of the spectrum which is evident from   fig.~\ref{fig:Lumperunitwavelength}.  The increase in mass shifts the spectrum towards a higher wavelength.  This results in a reduction in the X-ray band (the lower wavelength) luminosity but an enhancement in the IR band luminosity. The increase in mass also increases the peak luminosity which falls in the UV band for the SMBHs. On the other hand, the X-ray and UV part of the spectrum also shift rightwards as the black hole spins down whereas the Vis-IR part of the spectrum remains almost unaffected. Therefore the spin-down of the SMBHs reduces the X-ray and UV band luminosities more dominantly as compared to the Vis-IR band. Thus the distributions of AGNs in various activity planes are non-uniform.  The non-uniformity in the distribution of the Eddington ratio, which is independent of the mass of the black hole, is due to the non-linear dependence on the spin of the SMBH (eq.~\ref{eq:fedd}). 

The lowermost panel of  fig.~\ref{fig:Lfeddbands1e-19} shows the luminosity and $f_\textrm{Edd}$ distributions after $t=7\times 10^8$ years ($\approx t_\textrm{Edd}$). As superradiant instability spins these black holes down to the boundary of the depletion region, the distribution in the activity planes narrows down along the various tracks corresponding to the three modes. These tracks set specific relations  among  $L_{\mrm{IR}},~ L_{\mrm{UV}},~ L_{\mrm{X}},~   f_\textrm{Edd}$ in the activity plane dictated by the scalar mass. Additionally, we also observe depletion regions in the various activity planes of AGNs. Most prominently seen in the planes $L_{\mrm{X}}-L_{\mrm{IR}}$, $ L_{\mrm{X}}-L_{\mrm{UV}}$ and  in the plane of $f_\textrm{Edd}$ vs different band-luminosities.

We have derived this relation keeping the accretion rate parameter $\dot{m}=0.5$ to be the same for all the AGNs and constant over time. In a realistic scenario, the accretion rate parameter can be different for different AGNs and can have time dependence as well. This can alter the epoch of spin-down but the critical spin, having dependence only on the mass of the scalar particle and the mass of the SMBH, remains unaffected. With the variation in the accretion rate parameter, one would hence expect a scatter about these tracks in the activity planes. 

The salient feature hence obtained from these distributions is the tendency of the AGNs to align along the tracks corresponding to the boundaries of the depletion region. The boundaries, on the other hand, are fully determined by the mass of the scalar. Therefore,  the presence of a scalar particle with a given mass would make the AGNs hosting an SMBH lying in the \sra region follow the superradiance-modified activity relation, irrespective of their accretion history. 

\subsection{Inclusion of self-interaction of scalar field}
It is important to note that this paper focuses on a \textit{free} ultra-light scalar field. Generally, we expect these fields to exhibit self-interaction. References \cite{sr4Baryakhtar:2020gao, Branco:2023frw} demonstrate that if the self-interaction strength $\lambda$ for a scalar field with quartic coupling ($\sim \lambda \Phi^4$) is sufficiently small, the evolution of the superradiant cloud remains unaffected by this self-interaction. Thus, the depletion regions in the spin-mass plane of the black hole remain unchanged as long as the $\lambda$ remains below the threshold value $\lambda_\textrm{th}$, as  given by \cite{sr4Baryakhtar:2020gao},
\begin{eqnarray}     
        \lambda_\textrm{th} &\approx \textrm{max}[10^{-86} \left(\frac{\mu}{10^{-19}~\textrm{eV}}\right)^{3/2}\left(\frac{T_\textrm{BH}}{10^{10}~yr}\right)^{-\frac{1}{2}} \left(\frac{\alpha}{0.01}\right)^{-\frac{11}{2}}, \nonumber \\ 
        &2\times10^{-94}\left(\frac{\mu}{10^{-19}~\textrm{eV}}\right)^2\left(\frac{0.01}{\alpha}\right)^{-\frac{3}{2}} \left(\frac{\tilde{a}}{0.9}\right)^{-\frac{1}{2}} ],
    \end{eqnarray}
where $T_\textrm{BH}$ is the age of the BH. Thus, the validity of the AGN observables derived under the assumption of free scalars can also be extended to self-interacting scalars, provided $\lambda < \lambda_\textrm{th}$.

\begin{figure}
    \centering
    \includegraphics[width=\linewidth]{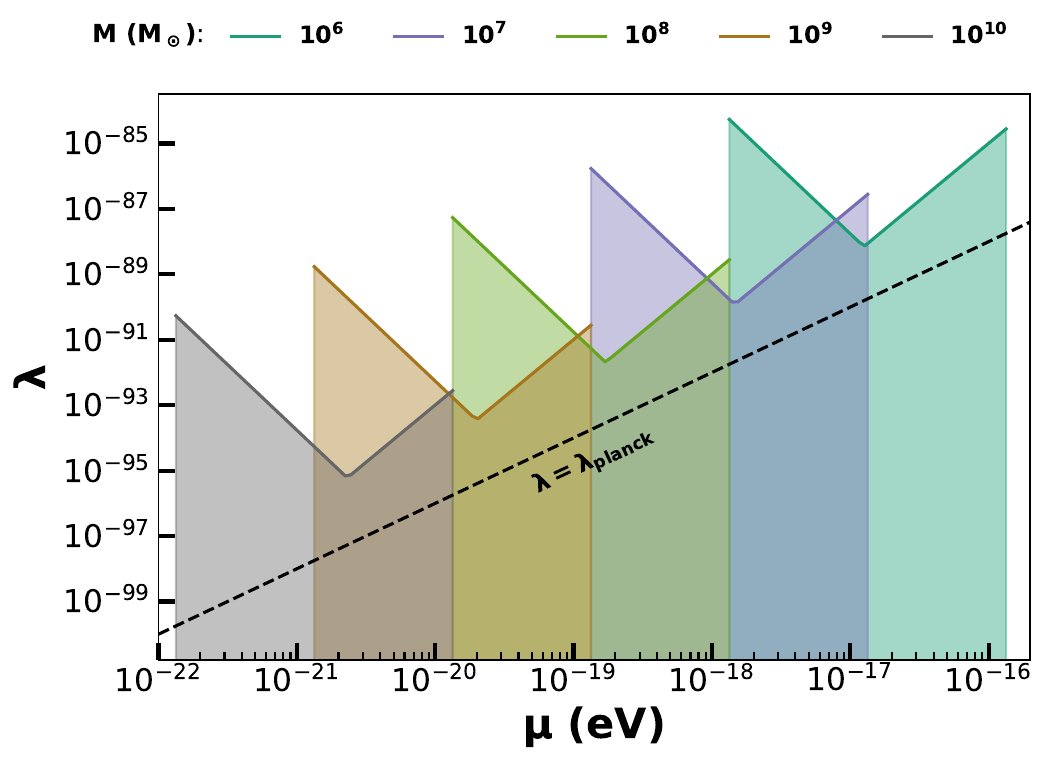}
    \caption{This figure illustrates the parameter space of the ultra-light scalar field with the quartic self-interaction strength $\lambda$, shown with a shaded region, which can be searched using the AGN characteristics proposed in this paper. The shaded region corresponds to the scalars with small enough self-interaction that the spin-down rate of SMBHs remains unaltered. Various colors of the shaded region correspond to different masses of SMBHs. Specifically, for axion-like particles with quartic coupling parametrized as $\lambda =\mu^2/f^2$, where $f$ is the symmetry breaking scale, the valid parameter region which can be probed using the AGN characteristics are the shaded region above the black-dashed line.}
    \label{fig:SelfInteraction}
\end{figure}

In fig.~\ref{fig:SelfInteraction}, we show shaded regions defined by the condition $\lambda<\lambda_\textrm{th}$. The shaded regions shown with different colors correspond to different masses of the SMBHs. Within these regions, our proposed time variation and the distribution of AGN characteristics can directly be utilized as a new probe for detecting the scalar field. However, care should be taken for interpreting the observables for large $\alpha>0.2$. As pointed out in ref.~\cite{sr4Baryakhtar:2020gao}, multiple levels beyond 011 and 022 are expected to grow in the presence of self-interaction, necessitating further investigation to fully understand the evolution.

In a specific model of scalar fields, when the self-interaction arises from the symmetry breaking at some energy scale $f_a$, also known as axion-like particles (ALPs), the $\lambda$ can be re-defined in terms of the symmetry breaking scale as $\lambda \equiv \mu^2/f_a^2$. For a valid description of ALPs, the maximum symmetry-breaking scale can reach up to the Planck scale ($\approx 10^{19}$~GeV). Consequently, we can define the minimum self-interaction for a valid description of ALPs and given by 
\begin{equation}
    \lambda_\textrm{planck} \approx 10^{-94}\left(\frac{\mu}{10^{-19}~\textrm{eV}}\right)^2.
\end{equation}
In the figure, we show a black dashed line representing $\lambda = \lambda_\textrm{planck}$, above which ALP description of the scalar field remains meaningful. Therefore, the parameter regime where AGN characteristics derived in this paper can be used to probe ALPs is defined by $\lambda_\textrm{planck}<\lambda<\lambda_\textrm{th}$ which corresponds to the area between the solid line boundary and the black dashed-line shown in the figure.

While this paper does not delve into the details of how self-interaction affects the evolution of scalar fields around accreting SMBHs in the parameter regime above the shaded region in the figure, we will briefly discuss the potential impacts and limitations of using AGN characteristics as observables for scalar fields, depending on the strength of $\lambda$ affecting the evolution as described in ref.~\cite{sr4Baryakhtar:2020gao}. It is worth mentioning here that as long as there is a significant spin-down of the SMBH, the AGN characteristics can, in principle, be utilized to search for the scalar field.

If $\lambda_\textrm{th}<\lambda<\lambda_\textrm{moderate}$, the $011$ mode grows fully but due to self-interaction early growth of the $022$ mode becomes possible, where $\lambda_\textrm{moderate}$ is given by~\citep{sr4Baryakhtar:2020gao}
\begin{eqnarray}
    \lambda_\textrm{moderate}&\approx 2\times10^{-89} \left(\frac{\mu}{10^{-19}~\textrm{eV}} \right)^2 \left(\frac{\tilde{a}(t_0)}{0.9}\right)^{-\frac{1}{2}} \nonumber\\
    & \textrm{max}[\left(\frac{\alpha}{0.04}\right)^{-\frac{3}{2}}, \left(\frac{\alpha}{0.04}\right)^{-3}].
\end{eqnarray}
This suggests that if the self-coupling is moderate, then it will prepone the spin-down of the SMBH due to the growth of the cloud in 022 mode. Hence, we will still be able to observe the drops in the luminosity of an AGN, except on the second drop corresponding to 022 mode with a smaller time scale. Thus, the depletion region can still be found in this parameter regime; it is just that the time scale of forming the depletion region corresponding to the 022 mode potentially be reduced.

For the self-coupling in the range $\lambda_\textrm{large}>\lambda>\lambda_\textrm{moderate}$, an equilibrium between the rate of superradiant production of the scalars and the rate of annihilation of the scalars to infinity will be achieved before the cloud in 011-mode saturates, where $\lambda_\textrm{large}$ is given by~\citep{sr4Baryakhtar:2020gao}
\begin{equation}
    \lambda_\textrm{large}\approx 10^{-91} \left(\frac{\mu}{10^{-19}~\textrm{eV}}\right)^3 \left(\frac{10^{10}~yr}{T_\textrm{BH}}\right)^{-1} \left(\frac{0.01}{\alpha}\right)^{-5} \left(\frac{0.9}{\tilde{a}(t_0)}\right)^{-\frac{3}{2}}.
\end{equation}
Hence, there will be a partial spin-down compared to the pure gravitational superradiance. Thus, depending upon the amount of change in the Luminosity of AGNs due to the partial spin-down, the observability of the effect will be reduced.

For a very large self-coupling $\lambda >\lambda_\textrm{large}$, there will not be any spin-down and hence the scalars can not be probed using AGN characteristics derived in this paper.

\section{Summary and discussion}\label{sec:conclusion}
An SMBH lying at the core of an AGN provides room for the elusive ULSPs to get produced through the phenomena of superradiance. Spin-down in these SMBHs, triggered by the superradiant instability, therefore naturally opens up a window to look for the  ULSPs in the mass range $10^{-20}$-$10^{-16}$~eV. In this work, we present a study of \srI experienced by the SMBH in the vicinity of AGN. We begin by showing in a realistic ambiance created by the accretion disk around the AGN, there is an enhanced growth of the scalar cloud and GW emission rate. We then discuss the effects of \sr on the characteristics of the AGN where we show the sudden drops in the time-variations of the Eddington ratio and various band-luminosities. Finally, we demonstrate the appearance of depleted regions and accumulations along the boundaries of those regions in the distribution of AGNs in the planes of band-luminosities and $f_{\mrm{Edd}}$. 

We start by numerically solving the time-evolution equations in  eq.~\ref{eq:EvolveWithAcc} of the scalar cloud along with the spin and mass of the SMBH in the presence of accretion. Throughout this study, we have taken a constant accretion rate parameter $\dot{m}=0.5$ which corresponds to the Eddington ratio spanning roughly from 0.03 to 0.15, depending upon the black hole's spin. This interval is consistent with the observed range $f_\textrm{Edd}\in(10^{-2},1)$ for AGNs as documented by SDSS~\citep{2020ApJS..249...17R}. One of the salient effects of accretion on \sr is the amplified growth in the mass of the cloud relative to the BH mass. The mass of the cloud is seen to increase by up to 25\% of the BH mass, which is in contrast with the case of no accretion where the expected growth is only up to 10\%. This enhancement occurs when the accretion rate is comparable to the spin-down rate due to superradiance. It is then the accretion that keeps the BH spin hover always above the critical spin $a_{\mrm{crit}}$ by feeding angular momentum to the BH  efficiently enough so that the superradiant instability remains switched on for a longer time.  This result from our time-evolution solution is in good agreement with the proposed ``over-superradiance" phase of the scalar cloud growth in ref.~\citep{sr9Hui:2022sri}. 

Another important outcome derived from this evolution is the possibility of obtaining modes of \sr with higher quantum numbers ($l,m=2,3$) within the lifetime of the universe.
As we observe from fig.~\ref{fig:Evolution}, this is achievable only because of the accretion phase that the SMBH goes through. The reason behind this is that the superradiant growth-time for the mode $nlm$  $\tau_{\mrm{sr}} \sim \frac{M}{\alpha^{4l+5}}$ reduces as the gravitational coupling $\alpha=M\mu$ tends to increase over time because of accretion, thus making it possible for the BH to pass through the 011-mode, 022-mode, and 033-mode dominated mass range consecutively within an observable time. Lastly, we notice that the growing scalar cloud ($M_s/M$) and increasing $\alpha$ have a remarkable effect on the power of GW emission rate which has been seen to increase by almost eight orders of magnitude than that of the non-accreting case.

With the solutions of the time-evolution of mass and spin of the SMBH, we derive the spin-dependent luminosities using the NT model of the accretion disk. We assume a quasi-static equilibrium of the accretion disk and argue for the validity of the assumption by comparing the timescale of \srI to the orbital period of a test particle near ISCO. We divide the full color-corrected spectrum into three wavelength bands-X-ray ($10^{-4}-10^{-2}~\mu$m), UV (0.010-0.4 $\mu$m), and Vis-IR (0.4-100~$\mu$m). In figs.~\ref{fig:EddingtonRatio} and \ref{fig:Lbands}, we observe sudden drops in the X-ray, UV, Vis-IR luminosities and the Eddington ratio with a characteristic time-scale of superradiant growth. In particular, one observes three drops in the time variation corresponding to the three modes of superradiance. Furthermore, the luminosity depends on the radiation efficiency $\epsilon(\ta)$ (eq.\ref{eq:Lbol}) which is a function of the BH spin. Therefore the amplitude of the drop in the luminosity can be characterized by the change in the spin of the BH due to superradiance. Similar observations regarding the instant drops can be made for the Eddington ratio $f_{\mrm{Edd}}$ which is also a function of $\epsilon(\ta)$. To know the exact spin-down effect on $f_{\mrm{Edd}}$, we have also shown how it varies when there is no superradiance. The Eddington ratio,  in the absence of superradiant instability, is expected to grow monotonically until it saturates at the maximum value in a relatively short time. In contrast to this monotonic growth, $f_{\mrm{Edd}}$ is seen to reach the maximum value at a late time after facing sudden drops due to the superradiant growth.  

In fig.~\ref{fig:Lfeddbands1e-19}, we present the distribution of various luminosity bands of AGNs in the presence of a scalar with mass $\mu=10^{-19}$ eV. Starting with a uniform distribution of $10^4$ SMBHs in the spin vs mass plane (Regge plane), we show their superradiant evolution in the presence of a scalar with mass $\in [10^{-20}$-$10^{-16}]$ eV and for constant $\dot{m}=0.5$ in fig.~\ref{fig:Regge_hist}. As the SR-active black holes start spinning down due to the growth of the scalar cloud in each of the three modes $011,022,033$,  one can infer  two visible effects -- the formation of gaps/depletion regions and accumulation of the SMBHs along a boundary of the depletion region in the Regge plane. For a given scalar mass, the boundary along which accumulation happens can be approximated by $\tilde{a}_\mrm{crit}(M,\mu, m)$  curve. The fate of the distribution of the BHs is determined by the timescales of \sr and accretion. The red-dashed curve represents $\tau_\textrm{growth}^{nlm}=\tau_\textrm{acc}$ above which there is a population of the SMBHs with $\tau_\textrm{growth}^{nlm} < \tau_{acc}$ where the black holes first spin down to the boundary/critical spin curve and then continue to move along the curve due to accretion. On the other hand, the population with $\tau_\textrm{growth}^{nlm} > \tau_{acc}$  remain below the boundary until accretion drags them into the depletion region and eventually leads them to satisfy the \sr condition. The consequence of these characteristics seen in the Regge plane can be mapped to the distribution in the various luminosity planes. In fig.~\ref{fig:Lfeddbands1e-19}, we see how the initial distribution in fig.~\ref{fig:InitialLfeddbands1e-19} of the AGNs on different planes of luminosities and Eddington ratio later evolves due to \sr caused by a scalar of mass $\mu=10^{-19}$eV. Here also we observe similar depletion regions and clustering along the various tracks where the tracks correspond to taking into account the three modes of superradiance. The depleted region is   visible on  the planes $L_{\mrm{X}}-L_{\mrm{IR}}$, $ L_{\mrm{X}}-L_{\mrm{UV}}$ and in the plane of $f_\textrm{Edd}$ vs different band-luminosities. As mentioned earlier, throughout the evolution, we assume the accretion rate parameter to be constant. 

The novel feature that is evident from these distributions therefore is the accumulation of AGNs along the tracks that correspond to the boundary of the depletion region which is determined by the mass of the scalar. Hence, the existence of a scalar field with a given mass would lead the AGNs hosting an SMBH lying in the \sra region to follow the superradiance-modified activity relation, irrespective of their accretion history. With the increased statistics and precision in the measured characteristics of the AGNs with experiments like DESI~\citep{DESI}, we may hope to observe the emergence of such accumulations of AGNs along the proposed superradiant activity tracks computed in this paper. To place robust constraints on these ultralight scalars from the various luminosity-mass relations, one would in practice have to search for these superradiance-modified relations that we show in this work. 

In practice,  $\dot{m}$ will also have some time-variation. As shown in fig.~\ref{fig:Depletion}, the depletion regions in both the Regge and the luminosity planes are expected to get reduced with a higher value of $\dot{m}$ for a fixed scalar mass. With lower values of $\mu$, one observes further reduction in the depletion region. Therefore, for smaller scalar mass and higher accretion rate, it may be possible that one would not find any depletion region in the spin versus mass plane. We would also like to point about one crucial phenomenon that arises while dealing with the black hole superradiance in a realistic environment like AGN- emission of jets \citep{Narayan:2021qfw}. Jet feedback, which leads to the extraction of angular momentum of BH, could be a contemporary process with BH superradiance, and the interplay between these two spin extraction mechanisms may affect the signal that we are showing in this work. However, \cite{Ricarte:2023owr} shows that for thin disk and $f_\textrm{Edd}$ roughly lying in the range $0.01-0.2$, the jet feedback is sub-dominant as compared to the accretion, and hence accretion will dominantly lead the BH to the extremal spin value. So, jet extraction can be ignored when considering $f_\textrm{Edd}$ within the range of 0.01 to 0.2, a common scenario in which we demonstrate the impacts of superradiance. However, for higher $f_\textrm{Edd}$ or even super-Eddington scenario ($f_\textrm{Edd}>1$), the presence of a jet can potentially alter the distribution of the AGN characteristics, and hence our proposed signature of superradiance may get diluted in this limit. 

We conclude this work by discussing a few potential observational signatures of the \sra AGN arising from the aspects of temporal changes in luminosities across different spectral bands of AGN caused by the superradiant spin-down of the central SMBH. We list the following potential observable phenomena at galactic centers that can carry the imprints of the superradiant history of the central SMBH in AGNs.
\begin{enumerate}
    \item A promising signature of \sr instability history of the SMBH at the center of an AGN can be found in the \ita{Galactic outflow} at the galactic center. The galactic outflow which is the massive depletion of gas from the central part of a galaxy, is a link that connects the center black hole to its host galaxy, and was observed in terms of massive gas outside radio quasars~\citep{GO5, GO6}.  Radiation-driven galactic outflows can carry information on superradiant instabilities, as such cases the outflow is quantified by the momentum transferred by radiation to the gas, which in turn depends on the luminosity ($L/c$). Therefore the sudden drops in the luminosity, particularly for the most luminous UV part of the AGN spectrum, that we find can occur due to superradiant instabilities (see fig.~\ref{fig:Lbands}),  can alter the expected radial profile of the galactic outflow. The profile of the galactic outflow depends on the time-scale and the amplitude of the change in the luminosities~\citep{GO1, GO2, GO3, GO4}. In the case of SR-activity of an AGN, the time-scale and the amplitude will be determined by the gravitational coupling of the scalar field with the black hole at the center. Thus, it can provide an observable to search for the superradiant history of the AGN so that even if the AGN is obscured and precise mass and spin measurements are not available, the characteristic outflow profile can give us a smoking gun signature of the superradiant instability of the ULSP. The observation of galactic outflow by the telescope such as the James Webb Space Telescope (JWST) may provide an opportunity to look for such signatures of superradiance~\citep{2023A&A...672A.128C, 2024ApJ...960..126V, 2024MNRAS.528.4976D}.

    \item Another interesting signature of the sudden luminosity drops in the UV band can appear in the Ly-$\alpha$ emission line and  Ly-$\alpha$ forest. Without the presence of a scalar that causes superradiance, one would expect that there will be continuous ionization of the neutral gas in the vicinity of a bright UV source. This effect is known as \ita{proximity effect}~\citep{proximityeff}. With this effect, since there will be a large depletion of neutral Hydrogen gas, absorption of the  Ly-$\alpha$ photons will be weaker, leading to a weakened  Ly-$\alpha$ forest. In the presence of superradiance, the rate at which gas was previously ionized would be lower because of sudden drops in the luminosity. The effect on the  Ly-$\alpha$ emission line can be understood as the rate at which the UV photons caused the Ly-$\alpha$ transition would now be reduced due to a fall in the luminosity. 

    \item Today, we may find many BHs with the same mass are having  ``anomalous'' spins (under the assumption that AGNs with similar mass will have similar accretion histories) due to superradiance. Presumably,  if the mass and the accretion history are the same then we would statistically expect a similar spin distribution. This example should be valid even if the accretion history depends significantly on the spin. There would be competition however with signatures of SMBH growth by chaotic accretion. This might lead to spin-down, as for superradiance, but also to flaring, offering a possible discriminant from the effects of superradiance.
    
    \item Chaotic accretion can grow the seed BH faster than accretion through disks. The reason is that the accretion disk-driven growth also increases the spin of the BH very fast which further increases the radiative efficiency and hence reduces the BH growth rate. Now, in the presence of a scalar field in the universe, light seed BH will never be able to go above a certain critical spin leading to suppression in $f_\textrm{Edd}$ as shown in fig.~\ref{fig:EddingtonRatio} and hence their growth can be large enough to produce the massive BHs. Hence, there can be a superradiant boost in the growth of seed BH due to their spin-down. Since accretion disk-driven growth also transforms the BH into a very luminous source, the observed characteristics of the massive AGNs may offer an opportunity to search for the presence of a scalar field in our universe.    
\end{enumerate}

\section*{Acknowledgements}
We appreciate the valuable comments provided by the anonymous referee on the manuscript. We would also like to thank Richard Brito and Martin Spinrath for their useful comments and suggestions on the manuscript. The work of PS and KC was supported by the National Science and Technology Council (NSTC) of Taiwan under grant no. MOST-110-2112-M- 007-017-MY3. PS and HV would like to express a special thanks to IAP, Paris for its hospitality and support during the visit. HV would like to thank the financial support from IIT Bombay to visit IAP. 

\section*{Data Availability}
The data backing the plots depicted in this article, as well as other findings from this study, will be made available upon reasonable request to the corresponding author.



\bibliographystyle{mnras}
\bibliography{BlackSupAGN} % if your bibtex file is called example.bib

% Don't change these lines
\bsp	% typesetting comment
\label{lastpage}
\end{document}